\documentclass[a4paper,fleqn,usenatbib]{mnras}

\usepackage{newtxtext,newtxmath}

\usepackage[T1]{fontenc}
\usepackage{ae,aecompl}

\usepackage{graphicx}	
\usepackage{amsmath}	
\usepackage{amssymb}	

\newcommand{\be}{\begin{eqnarray}}
\newcommand{\ee}{\end{eqnarray}}

\title[Impact of the reflection model]{Impact of the reflection model on the estimate of the properties of accreting black holes}

\author[Tripathi, Liu \& Bambi]{
Ashutosh~Tripathi, Honghui Liu and Cosimo~Bambi\thanks{Corresponding author: bambi@fudan.edu.cn}
\\
Center for Field Theory and Particle Physics and Department of Physics, Fudan University, 200438 Shanghai, China}

\begin{document}
\label{firstpage}
\pagerange{\pageref{firstpage}--\pageref{lastpage}}
\maketitle

\begin{abstract}
Relativistic reflection features in the X-ray spectra of black hole binaries and AGNs originate from illumination of the inner part of the accretion disk by a hot corona. In the presence of high quality data and with the correct astrophysical model, X-ray reflection spectroscopy can be quite a powerful tool to probe the strong gravity region, study the morphology of the accreting matter, measure black hole spins, and even test Einstein's theory of general relativity in the strong field regime. There are a few relativistic reflection models available today and developed by different groups. All these models present some differences and have a number of simplifications introducing systematic uncertainties. The question is whether different models provide different measurements of the properties of black holes and how to arrive at a common model for the whole X-ray astronomy community. In this paper, we start exploring this issue by analyzing a \textsl{Suzaku} observation of the stellar-mass black hole in GRS~1915+105 and simultaneous \textsl{XMM-Newton} and \textsl{NuSTAR} observations of the supermassive black hole in MCG--6--30--15. The relativistic reflection component of these sources is fitted with {\sc relconv$\times$reflionx}, {\sc relconv$\times$xillver}, and {\sc relxill}. We discuss the differences and the impact on the study of accreting black holes.
\end{abstract}




\section{Introduction}

Relativistic reflection features are commonly observed in the X-ray spectra of black hole binaries and AGNs~\citep{1989MNRAS.238..729F,1995Natur.375..659T,2007MNRAS.382..194N,2013MNRAS.428.2901W,2014SSRv..183..277R,2018AnP...53000430B}. For an Eddington-scaled disk luminosity of $\sim$10\% and a large angular momentum of the accreting gas, a black hole is surrounded by a geometrically thin and optically thick accretion disk. The gas in the disk is in local thermal equilibrium, every point of the disk has a blackbody-like spectrum, and the whole disk has a multi-temperature blackbody-like spectrum~\citep{1973A&A....24..337S,1974ApJ...191..499P}. The thermal spectrum of the accretion disk is peaked in the soft X-ray band for stellar-mass black holes and in the optical/UV bands for supermassive black holes. The ``corona'' is some hotter ($\sim$100~keV), often compact and optically thin, electron cloud near the black hole~\citep{1979ApJ...229..318G}. Thermal photons from the disk can inverse Compton scatter off free electrons in the corona, generating a power-law spectrum with a high energy cut-off~\citep{1979Natur.279..506S}. Comptonized photons can illuminate the accretion disk, producing a reflection spectrum~\citep{1991MNRAS.249..352G}. The latter is characterized by some fluorescent emission lines, notably the iron K$\alpha$ complex in the soft X-ray band, and the Compton hump peaked at 20-30~keV.
The reflection features in the X-ray spectra of black holes are affected by the accretion disk geometry and the strong gravitational field around the compact object~\citep{1989MNRAS.238..729F,1991ApJ...376...90L}. In the presence of high quality data and with the correct astrophysical model, the analysis of these reflection features can be used to study accreting black holes, investigate the properties of the accretion disks, measure black hole spins, and even test Einstein's theory of general relativity in the strong field regime~\citep{2014SSRv..183..277R,2017RvMP...89b5001B}.

In the past 5-10 years, there has been significant progress in the development of the analysis of these features, thanks to both more sophisticated astrophysical models and new observational data. Today there are a few available reflection models that have been developed by different groups. While these models are the state of the art in the field, they are based on a number of simplifications that introduce systematic uncertainties in the final measurements of the parameters of the system. Depending on the quality of the data, such systematic uncertainties may be negligible or lead to biased estimates of the properties of accreting black holes.

Systematic uncertainties related to the theoretical model\footnote{Here we ignore systematic uncertainties of other origin, such as those caused by instrumental effects (calibration and energy resolution) or data analysis softwares.} can be grouped into four classes:
\begin{enumerate}
\item {\it Accretion disk}. The accretion disk is normally described by the Novikov-Thorne model~\citep{1973blho.conf..343N,1974ApJ...191..499P} (Keplerian disk on the plane perpendicular to the black hole spin) and approximated as infinitesimally thin. The inner edge of the disk is at the innermost stable circular orbit (ISCO) or at a larger radius, and there is no emission from the plunging region, namely the region between the black hole and the inner edge of the disk. However, a real accretion disk has a finite thickness and some radiation is also emitted from the plunging gas~\citep[see, for instance,][]{2020MNRAS.493.5389F}, two effects that have some impact on the estimate of the model parameters~\citep{2008ApJ...675.1048R,2018ApJ...855..120T,2020MNRAS.493.5532W,2020arXiv200506719C}. Moreover, it is common to use these thin disk reflection models even to fit the spectra of sources with high mass accretion rate, and this can lead to a significant bias in the estimates of some properties of the system~\citep{2019arXiv191106605R,2020MNRAS.491..417R}. Most models assume that the density and the ionization parameter are constant over the radial coordinate of the disk, but this has been argued to produce modeling bias in the estimate of some parameters~\citep{2012A&A...545A.106S,2019MNRAS.485..239K} and thus some of the most recent models include the possibility for ionization gradients with some prescription for the density profile~\citep[see, for instance,][]{2019MNRAS.488..324I}.
\item {\it Corona}. The morphology of the corona is not well understood as of now, and it is likely that it changes as the source moves to different spectral states~\citep{2015MNRAS.449..129W}. The coronal geometry determines the emissivity profile of the reflection spectrum over the disk, and the choice of the emissivity profile model can have quite remarkable impacts on the measurements of the parameters of the system~\citep{2019ApJ...884..147Z}. The emissivity profile from coronae of arbitrary geometry are normally modeled with a simple power-law or a broken power-law. These are clearly approximations that can introduce important biases in the presence of high quality data. For coronae with specific geometry, the most popular choice is the lamppost set-up, where the corona is approximated as a point-like source along the black hole spin axis~\citep{2013MNRAS.430.1694D}. There are only a few studies of coronae with more complicated geometries~\citep{2013MNRAS.430.1694D,2012MNRAS.424.1284W}.  
\item {\it Reflection spectrum}. Illumination of the accretion disk by a hot corona generates the reflection spectrum. Calculations of the reflection spectrum in the rest-frame of the gas in the disk involves atomic physics and all the available models assume a number of simplifications that introduce systematic uncertainties on the final measurements of the system. Current models assume a cold accretion disk. {\sc reflionx} and {\sc xillver} carry a 1D calculation assuming a plane-parallel slab, and the density is assumed to be constant in the vertical direction \citep[see, however,][where the density in the vertical direction is self-consistently calculated]{2004ApJ...603..436B,2002MNRAS.332..799R}. The electron density of the angle-resolved reflection model {\sc relxilld} is allowed to vary between $10^{15}$~cm$^{-3}$ to $10^{19}$~cm$^{-3}$~\citep{2016MNRAS.462..751G}, while higher densities, which would be expected in the disks of black hole binaries, are possible with the angle-averaged model {\sc reflionx}~\citep{2018ApJ...855....3T,2019MNRAS.484.1972J,2019MNRAS.489.3436J}. Photons from the corona illuminate the disk at a fixed inclination angle. Only the iron abundance is free to vary during the fitting, while the abundances for all other elements are fixed to their solar values.
\item {\it Relativistic effects}. Some relativistic effects occurring in the strong gravity region are ignored even by the current more sophisticated reflection models. Returning radiation, namely the radiation emitted by the disk and returning to the disk because of the strong light bending near the black hole, is ignored in the available models and its effect may not be negligible~\citep{2018MNRAS.477.4269N,2020arXiv200615838R}. Higher order disk images, namely radiation emitted by the disk and circling the black hole one or more times, is ignored but its impact is also very weak~\citep{2020PhRvD.101d3010Z}. The exact coronal spectrum illuminating the disk is determined by the gravitational redshift between the location of the corona and the point on the disk, and it can only be calculated if the coronal geometry is known.
\end{enumerate}
The next generation of X-ray missions (e.g. \textsl{XRISM}, \textsl{eXTP}, \textsl{Athena}, \textsl{STROBE-X}) promises to provide unprecedented high quality data, which will necessarily require more accurate synthetic reflection spectra than those available today and a better knowledge of the physical properties of these systems.

In this paper we start exploring the issue of whether the current reflection models provide different measurements of the parameters of accreting black holes and how to arrive at a common model for the whole X-ray astronomy community. As a first step to this direction, we analyze some high quality data by fitting the reflection components of two sources with different models. We consider a 117~ks observation of \textsl{Suzaku} of the black hole binary GRS~1915+105 and a set of simultaneous observations with \textsl{XMM-Newton} and \textsl{NuSTAR} of the AGN MCG--6--30--15.

We chose these sources/data to have a black hole binary and an AGN, one observed from a high viewing angle and the other one from a low viewing angle. As for GRS~1915+105, \textsl{Suzaku} provides both a good energy resolution near the iron line and data up to $\sim 50$~keV to see the Compton hump. That particular \textsl{Suzaku} observation shows a simple spectrum with strong relativistic reflection features and no disk thermal component, so the disk temperature was presumably low and this is consistent with the assumption of cold disk used in the reflection models. We are not aware of any \textsl{NuSTAR} observation of GRS~1915+105 with such properties. In the case of MCG--6--30--15, the source is a very bright AGN with a prominent and broad iron line, so it is quite a common source for testing reflection models. Here we have a good energy resolution from the \textsl{XMM-Newton} data and we can observe the Compton hump with \textsl{NuSTAR}.

We employ the reflection models {\sc relconv$\times$reflionx}, {\sc relconv$\times$xillver}, and {\sc relxill}. The three models provide the same measurements for GRS~1915+105 while show some discrepancy for MCG--6--30--15. We discuss the origin of the discrepancies and the impact on the analysis of future observations. We stress that, considering the complexity of current reflection models, it is not an easy task to understand their systematic uncertainties. The choices of the sources, observations, and models may combine in a quite complicated way and it is difficult to arrive at very general conclusions on these reflection models. Additional work is thus necessary for a complete comprehension of this issue.

The content of this paper is as follows. In Section~\ref{s-models}, we review the reflection models employed in our study, pointing out their differences. In Section~\ref{s-grs}, we present the analysis of a 2012 \textsl{Suzaku} observation of the black hole binary GRS~1915+105 and we fit the relativistic reflection component with {\sc relconv$\times$reflionx}, {\sc relconv$\times$xillver}, and {\sc relxill}. In Section~\ref{s-mcg}, we present a similar analysis of some simultaneous observations with \textsl{XMM-Newton} and \textsl{NuSTAR} of the AGN MCG--6--30--15. We discuss our results in Section~\ref{s-dc}.


\section{Relativistic reflection models}\label{s-models}

The calculations of the reflection spectrum of the accretion disk of a black hole are normally split into two parts: $i)$ the calculations of the reflection spectrum in the rest-frame of the gas, and $ii)$ the calculations of the spectrum at the detection point far from the source for a given spectrum at the emission point on the accretion disk. Part $i)$ depends on atomic physics (particle scattering, atomic energy levels, elemental abundance, etc.) and on some assumption about the material of the disk (e.g. density profile in the vertical direction), while does not directly depend on the actual geometry of the whole disk and the spacetime metric. In other words, only micro-physics is involved. Part $ii)$ concerns the macro-physics: the geometry of the accretion disk, the disk emissivity profile (which depends on the coronal geometry), and the spacetime metric (which affects the motion of the gas in the disk and determines all the relativistic effects).

Calculations of parts $i)$ and $ii)$ are connected employing the formalism of the transfer function~\citep{1975ApJ...202..788C,1995CoPhC..88..109S,2017ApJ...842...76B}. The observed reflection spectrum can be written as
\be
F_{\rm o} (\nu_{\rm o}) = \int g^3 I_{\rm e} ( r_{\rm e} , \vartheta_{\rm e} ) \, dX dY \, ,
\ee 
where $g = \nu_{\rm o}/\nu_{\rm e}$ is the redshift factor, $\nu_{\rm o}$ and $\nu_{\rm e}$ are, respectively, the photon frequency at the detection point and at the emission point in the rest-frame of the gas, $I_{\rm e}$ is the specific intensity of the radiation at the emission point in the rest-frame of the gas, $r_{\rm e}$ is the emission radius, $\vartheta_{\rm e}$ is the emission angle (i.e. the angle between the photon direction and the disk normal at the emission point in the rest-frame of the gas), and $X$ and $Y$ are the Cartesian coordinates on the observer's plane. Introducing the transfer function, the observed reflection spectrum can be written as  
\be\label{eq-fff0}
F_{\rm o} (\nu_{\rm o}) = \frac{1}{D^2} \int_{R_{\rm in}}^{R_{\rm out}} \int_0^1
\frac{\pi r_{\rm e} g^2 f(g^*, r_{\rm e} , i)}{\sqrt{g^* (1 - g^*)}} \, I_{\rm e} 
( r_{\rm e} , \vartheta_{\rm e} ) \, dg^* dr_{\rm e} \, ,
\ee
where $R_{\rm in}$ and $R_{\rm out}$ are, respectively, the inner and outer edges of the accretion disk, $g^*$ ranges from 0 to 1 and is the relative redshift factor defined as
\be
g^* = \frac{g - g_{\rm min}}{g_{\rm max} - g_{\rm min}} \, ,
\ee
where $g_{\rm max} = g_{\rm max} (r_{\rm e}, i)$ and $g_{\rm min} = g_{\rm min} (r_{\rm e}, i)$ are, respectively, the maximum and minimum redshift for photons emitted from the radius $r_{\rm e}$ and a disk's inclination angle $i$ (i.e. the angle between the normal to the disk and the line of sight of the distant observer). The transfer function is
\be
f(g^*, r_{\rm e} , i) = \frac{g \sqrt{g^* (1 - g^*)}}{\pi r_{\rm e}} \left| \frac{\partial \left(X,Y\right)}{\partial \left(g^*,r_{\rm e}\right)} \right| \, ,
\ee
where $| \partial (X,Y)/\partial (g^*,r_{\rm e}) |$ is the Jacobian. For infinitesimally thin disks in Kerr spacetimes, for given values of $r_{\rm e}$ and $i$ the transfer function is a closed curve parametrized by $g^*$. There is only one point with $g^* = 0$ and only one point with $g^* = 1$. There are two curves connecting these two points and there are thus two branches of the transfer function, say $f^{(1)}(g^*, r_{\rm e} , i)$ and $f^{(2)}(g^*, r_{\rm e} , i)$. We can thus rewrite $F$ as
\be\label{eq-fff}
\hspace{-0.5cm}
F_{\rm o} (\nu_{\rm o}) =& \frac{1}{D^2} \int_{R_{\rm in}}^{R_{\rm out}} \int_0^1
\frac{\pi r_{\rm e} g^2 f^{(1)}(g^*, r_{\rm e} , i)}{\sqrt{g^* (1 - g^*)}} \, I_{\rm e} 
( r_{\rm e} , \vartheta_{\rm e}^{(1)} ) \, dg^* dr_{\rm e} \nonumber\\
=& \frac{1}{D^2} \int_{R_{\rm in}}^{R_{\rm out}} \int_0^1
\frac{\pi r_{\rm e} g^2 f^{(2)}(g^*, r_{\rm e} , i)}{\sqrt{g^* (1 - g^*)}} \, I_{\rm e} 
( r_{\rm e} , \vartheta_{\rm e}^{(2)} ) \, dg^* dr_{\rm e} \, . 
\ee
Note that $\vartheta_{\rm e}^{(1)}$ and $\vartheta_{\rm e}^{(2)}$ are the emission angles in the branches 1 and 2. In the general case of non-isotropic emission, the local specific intensity $I_{\rm e}$ depends on the photon emission angle and this requires to calculate the two integrals separately. For isotropic emission, Eq.~(\ref{eq-fff}) reduces to Eq.~(\ref{eq-fff0}) with $f = f^{(1)} + f^{(2)}$. Note that in general $i \neq\vartheta_{\rm e}^{(1)} \neq \vartheta_{\rm e}^{(2)}$ because of the light bending of the spacetime.

In the rest of the paper, we will employ the two most popular models for the calculation of the reflection spectrum at the emission point in the rest-frame of the gas, namely {\sc reflionx}~\citep{2005MNRAS.358..211R} and {\sc xillver}~\citep{2010ApJ...718..695G,2013ApJ...768..146G}. These two models are used to calculate the shape of the local specific intensity $I_{\rm e}$, while the normalization of the spectrum is determined by the emissivity profile of the disk, which should depend on the coronal geometry. We note that {\sc xillver} provides an angle-resolved reflected flux, while {\sc reflionx} only provides the angle-averaged reflection intensity. {\sc xillver} also incorporates a richer atomic database than {\sc reflionx}~\citep[see][]{2013ApJ...768..146G}. As a convolution model, we will use {\sc relline}/{\sc relconv}~\citep{2010MNRAS.409.1534D,2013MNRAS.430.1694D}, in which the disk is described by an infinitesimally thin Novikov-Thorne disk and the spacetime is described by the Kerr metric. We will thus consider three possible relativistic reflection models: {\sc relconv$\times$reflionx}, {\sc relconv$\times$xillver}, and {\sc relxill}. {\sc relconv$\times$reflionx} and {\sc relconv$\times$xillver} are, respectively, the reflection spectra calculated with {\sc reflionx} and {\sc xillver} and convolved with {\sc relconv}. The two models only differ from the calculation of the reflection spectrum in the rest-frame of the gas and employ the same convolution model. In particular, both models assume that $i = \vartheta_{\rm e}^{(1)} = \vartheta_{\rm e}^{(2)}$, which is an approximation and does not take the position-dependent photon emission angle into account. 
Indeed, because of light bending $\vartheta_{\rm e}^{(1)}$ and $\vartheta_{\rm e}^{(2)}$ should change at every point on the disk and be different from $i$, but when we apply {\sc relconv} to {\sc reflionx} or {\sc xillver} we ignore such a difference (angle-averaged models). {\sc relxill}~\citep{2013MNRAS.430.1694D,2014ApJ...782...76G} employs the reflection model {\sc xillver} and is equivalent to {\sc relconv$\times$xillver} but without the assumption $i = \vartheta_{\rm e}^{(1)} = \vartheta_{\rm e}^{(2)}$ (angle-resolved model). Note that such an approximation is justified by the fact that relativistic effects (Doppler boosting) are very sensitive to the inclination angle of the disk $i$, while the local specific intensity $I_{\rm e}$ is not so much sensitive to the exact photon emission angle $\vartheta_{\rm e}$.


\begin{figure}
\begin{center}
\includegraphics[width=8.0cm,clip]{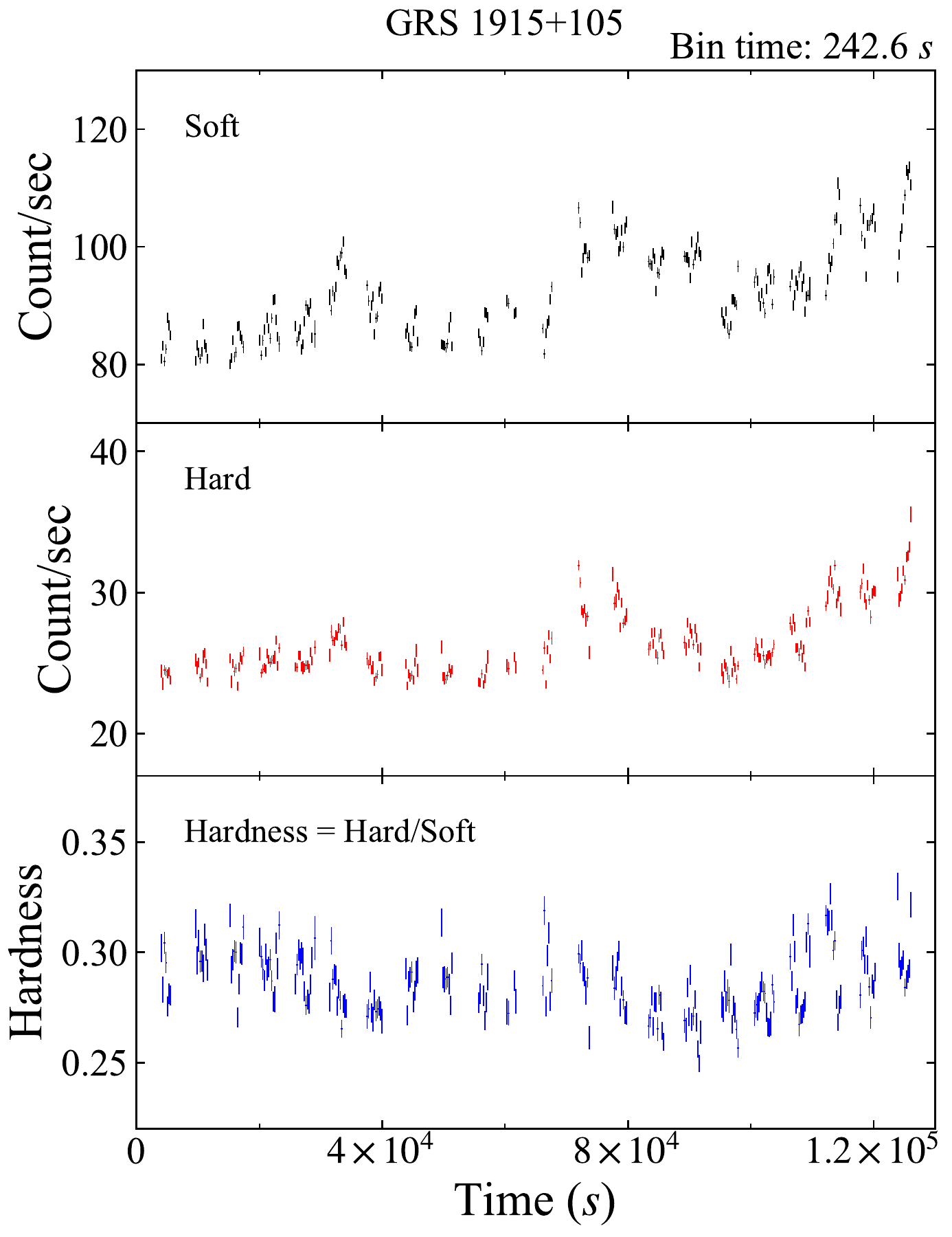}
\end{center}
\vspace{-0.2cm}
\caption{GRS~1915+105 -- Light curves of the \textsl{Suzaku}/XIS1 data (0.2-12~keV, upper panel), \textsl{Suzaku}/HXD data (10-70~keV, central panel), and temporal evolution of the hardness of the spectrum (bottom panel). From~\citet{2019ApJ...884..147Z}. \label{f-grs-lc}}
\begin{center}
\includegraphics[width=8.5cm,trim={2.6cm 0cm 2.6cm 18.2cm},clip]{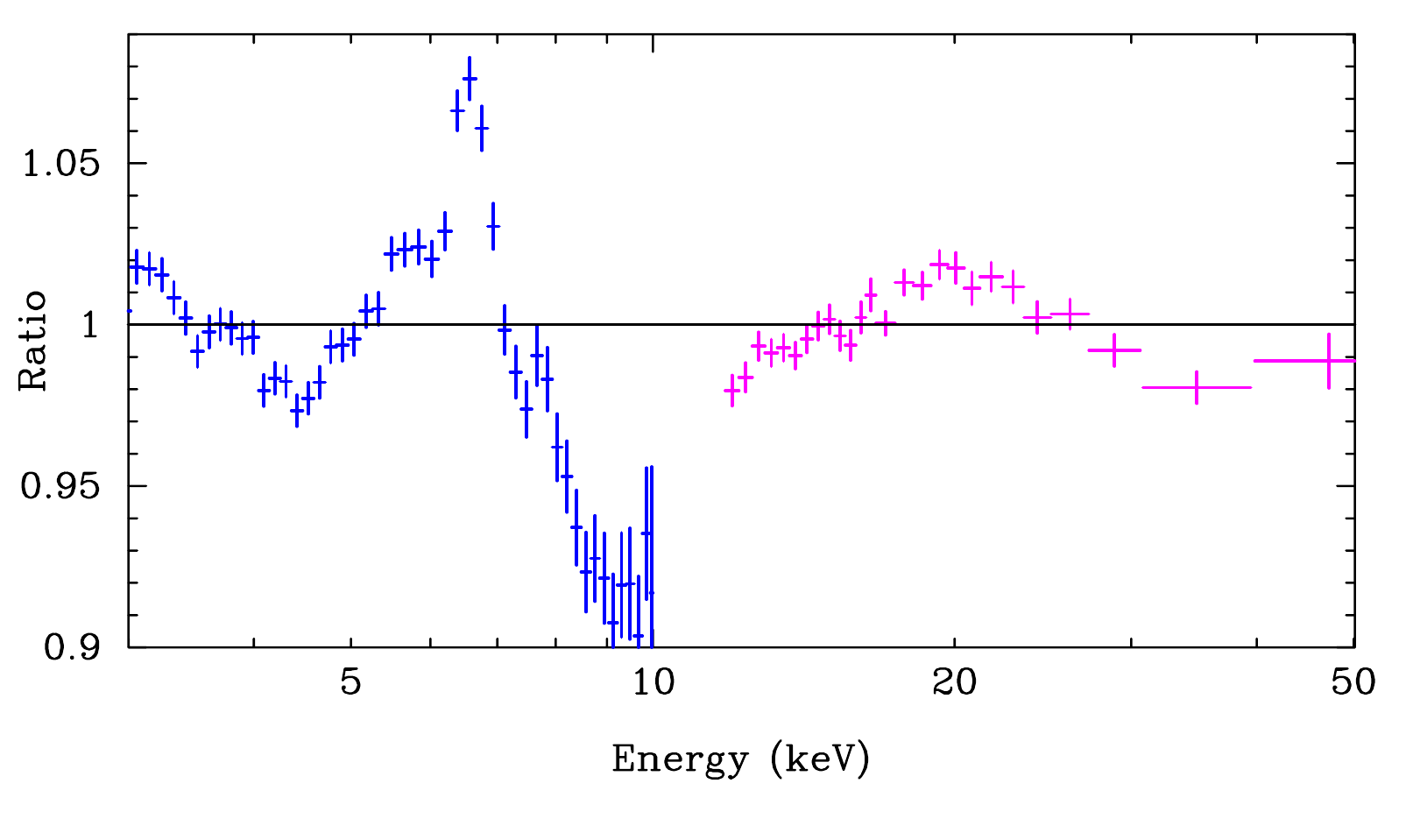}
\end{center}
\vspace{-0.4cm}
\caption{GRS~1915+105 -- Data to best-fit model ratio for {\sc tbabs$\times$cutoffpl}. Blue crosses are used for \textsl{Suzaku}/XIS1 data and magenta crosses are used for \textsl{Suzaku}/HXD data. \label{f-grs-ratio}}
\end{figure}

\section{Reflection spectrum of GRS~1915+105}\label{s-grs}

GRS~1915+105 is a black hole binary with some peculiar properties. While it is a low-mass X-ray binary, it is a persistent X-ray source since 1992, when it went into outburst and was discovered by the WATCH instrument on board of the \textsl{Granat} satellite~\citep{1992IAUC.5590....2C}. The most recent estimates of the source distance and black hole mass are, respectively, $D = 8.6_{-1.6}^{+2.0}$~kpc and $M = 12.4_{-1.8}^{+2.0}$~$M_\odot$~\citep{2014ApJ...796....2R}. GRS~1915+105 is a very variable source across all bands of the electromagnetic spectrum~\citep{2000A&A...355..271B} and is often considered as the archetype of microquasars~\citep{1999ARA&A..37..409M}. The black hole spin has been estimated either using the continuum-fitting method~\citep{2006ApJ...652..518M} and X-ray reflection spectroscopy~\citep{2009ApJ...706...60B,2013ApJ...775L..45M,2019ApJ...884..147Z,2020arXiv200309663A}, and all studies agree that the black hole is rotating very fast, with a spin parameter $a_*$ close to 1.

\begin{table}
\centering
{\renewcommand{\arraystretch}{1.25}
\begin{tabular}{l|c|c|c}
\hline\hline
Source & Satellite & Obs. ID & Total counts ($10^6$)\\
\hline
GRS~1915+105 & \textsl{Suzaku} & 402071010 & 2.43 (XIS1) \\
&&& 1.36 (HXD/PIN) \\
\hline
MCG--6--30--15 & \textsl{XMM-Newton} & 0693781201 & 1.74 (EPIC-Pn) \\
&& 0693781301 & \\
&& 0693781401 & \\
& \textsl{NuSTAR} & 60001047002 & 0.147 (FPMA) \\
&& 60001047003 & 0.141 (FPMB) \\
&& 60001047005 & \\
\hline\hline
\end{tabular}
\caption{Summary of the observations analyzed in the present work. The total counts refer to the energy range used in the analysis and specified in the text for every instrument. Note that in the case of MCG--6--30--15 the data are rearranged in four flux states and the total counts refer to the sum of all flux states.  \label{t-obsid}}
}
\end{table}

\subsection{Observation and data reduction}

\textsl{Suzaku} observed GRS~1915+105 on 2007-5-7 for approximately 117~ks (obs. ID~402071010), see Tab.~\ref{t-obsid}, while the source was in a low/hard flux state. The first analysis of this observation was reported in \citet{2009ApJ...706...60B}, where the authors find $a_* = 0.98 \pm 0.01$ (1-$\sigma$ confidence) in the analysis of the full broadband spectrum. As in \citet{2009ApJ...706...60B}, here we only use the data from the XIS1 and HXD/PIN instruments. Two other XIS units were turned off during the observation to preserve telemetry. The fourth unit was run in a special timing mode.

For the data reduction of both instruments, we follow \citet{2009ApJ...706...60B} and here we only report the main steps. Unfiltered event files of XIS1 are processed following the Suzaku Darta Reduction ABC Guide with \texttt{aepipeline} to create a clean event file using XIS CALDB version 20151005. Since XIS1 data are affected by pile up, the source region is an annulus (inner radius 78~arcsec, outer radius 208~arcsec). For the background region, we select an annulus with inner radius 208~arcsec and outer radius 278~arcsec. Redistribution matrix file and ancillary response file are created using the tools \texttt{xisrmfgen} and \texttt{xissimarfgen} available in the HEASOFT version~6.26.1 data reduction package. After all efficiencies and screening, the net exposure time is around 29~ks. Data are grouped using \texttt{grppha} to a minimum of 25~counts per bin. We only analyze the 2.3-10.0~keV energy band to avoid calibration problems near the Si~K~edge and because of the low number of photons at low energies due to the high absorption. For the HXD/PIN data, we proceed in a similar way. We use \texttt{aepipeline} and \texttt{hxdpinxbpi} with CALDB version 20110913. After all efficiencies and screening, the net exposure time is around 53~ks. Data are grouped using \texttt{grppha} to a minimum of 25~counts per bin. We analyze the 12.0-55.0~keV energy band. The cross-calibration constant between XIS1 and HXD/PIN data is left free, because the XIS data are affected by pile-up.

While GRS~1915+105 is normally a very variable source, its hardness was quite stable during this \textsl{Suzaku} observation, so the spectral analysis does not require any particular treatment to take the source variability into account. The light curves of the XIS1 and HXD/PIN data and the temporal evolution of the hardness of the source are shown in Fig.~\ref{f-grs-lc}.


\begin{table*}
\centering
{\renewcommand{\arraystretch}{1.25}
\begin{tabular}{l|c|c|c}
\hline\hline
Model & {\sc relconv$\times$reflionx} & {\sc relconv$\times$xillver} & {\sc relxill} \\
\hline
{\sc tbabs} &&& \\
$N_{\rm H} / 10^{22}$ cm$^{-2}$ & $5.16_{-0.03}^{+0.10}$ & $5.25_{-0.03}^{+0.03}$ & $5.19_{-0.03}^{+0.07}$ \\
\hline
{\sc cutoffpl} &&& \\
$\Gamma$ & $2.145_{-0.003}^{+0.027}$ & $2.200_{-0.017}^{+0.018}$ & $2.197_{-0.003}^{+0.009}$ \\
$E_{\rm cut}$ [keV] & $58.7_{-1.4}^{+0.8}$ & $71.7_{-6.0}^{+1.6}$ & $70.7_{-4.0}^{+1.7}$ \\
$N_\text{\sc cutoffpl}$ & $3.99_{-0.09}^{+0.10}$ & $3.93_{-0.15}^{+0.36}$ & $3.82_{-0.07}^{+0.29}$ \\
\hline
$q_{\rm in}$ & $10.00_{-0.24}$ & $9.9_{-0.3}^{\rm +(P)}$ & $10.0_{-1.6}$ \\
$q_{\rm out}$ & $0.0^{+0.3}$ & $0.00^{+0.22}$ & $0.00^{+0.21}$ \\
$R_{\rm br}$ [$M$] & $5.84_{-0.16}^{+0.47}$ & $6.3_{-0.3}^{+0.3}$ & $6.27_{-0.13}^{+0.47}$ \\
$a_*$ & $0.9945_{-0.0048}^{+0.0013}$ & $0.991_{-0.003}^{+0.004}$ & $0.9908_{-0.0017}^{+0.0015}$ \\
$i$ [deg] & $77.1_{-1.9}^{+0.6}$ & $73.5_{-1.9}^{+1.0}$ & $73.8_{-0.4}^{+0.8}$ \\
$\log\xi$ & $2.80_{-0.03}^{+0.03}$ & $2.74_{-0.05}^{+0.03}$ & $2.77_{-0.03}^{+0.05}$ \\
$A_{\rm Fe}$ & $0.46_{-0.03}^{+0.03}$ & $0.51_{\rm -(P)}^{+0.11}$ & $0.57_{-0.05}^{+0.07}$ \\
$N_\text{\sc reflionx}$ & $90_{-5}^{+4}$ & -- & -- \\
$N_\text{\sc xillver}$~$(10^{-3})$ & -- & $156_{-21}^{+10}$ & -- \\
$N_\text{\sc relxill}$~$(10^{-3})$ & -- & -- & $20.9_{-1.2}^{+0.5}$ \\ 
\hline
$\chi^2$/dof & $2350.55/2210 = 1.06360$ & $2335.56/2210 = 1.05682$ & $2328.70/2210 = 1.05371$ \\
\hline\hline
\end{tabular}
\vspace{0.2cm}
\caption{GRS~1915+105 -- Summary of the best-fit values for the models {\sc relconv$\times$reflionx}, {\sc relconv$\times$xillver}, and {\sc relxill}. The reported uncertainties correspond to the 90\% confidence level for one relevant parameter ($\Delta\chi^2 = 2.71$). $\xi$ in units of erg~cm~s$^{-1}$. In {\sc reflionx}, the fitting parameter is $\xi$, not $\log\xi$ as in {\sc xillver} and {\sc relxill}, but it is converted into $\log\xi$ in the table in order to facilitate the comparison with the other models. See the text for the details. \label{t-grs}}
}
\end{table*}

\begin{figure*}
\begin{center}
\includegraphics[width=5.7cm,trim={1.7cm 0cm 2.6cm 17.5cm},clip]{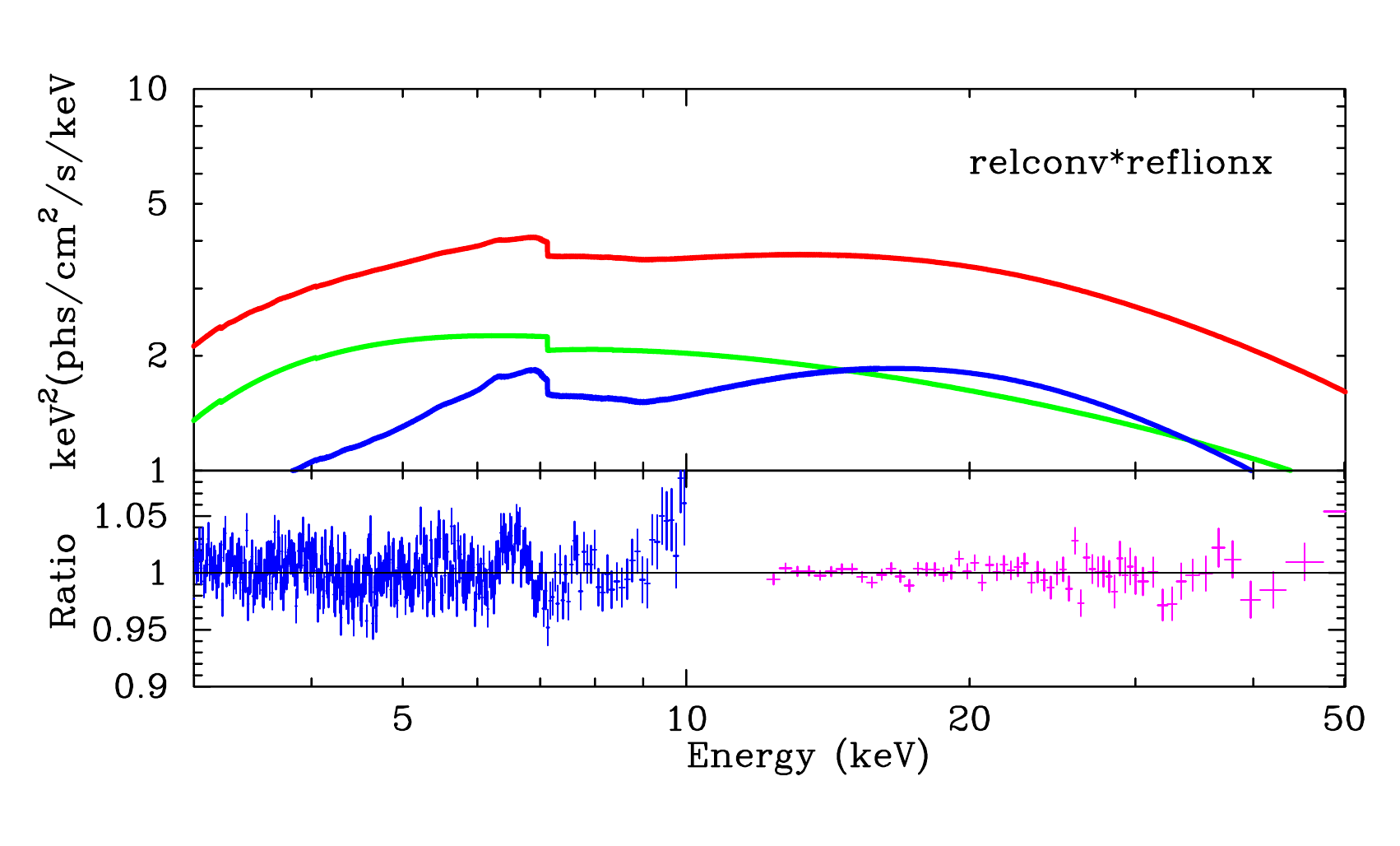}
\hspace{0.0cm}
\includegraphics[width=5.7cm,trim={1.7cm 0cm 2.6cm 17.5cm},clip]{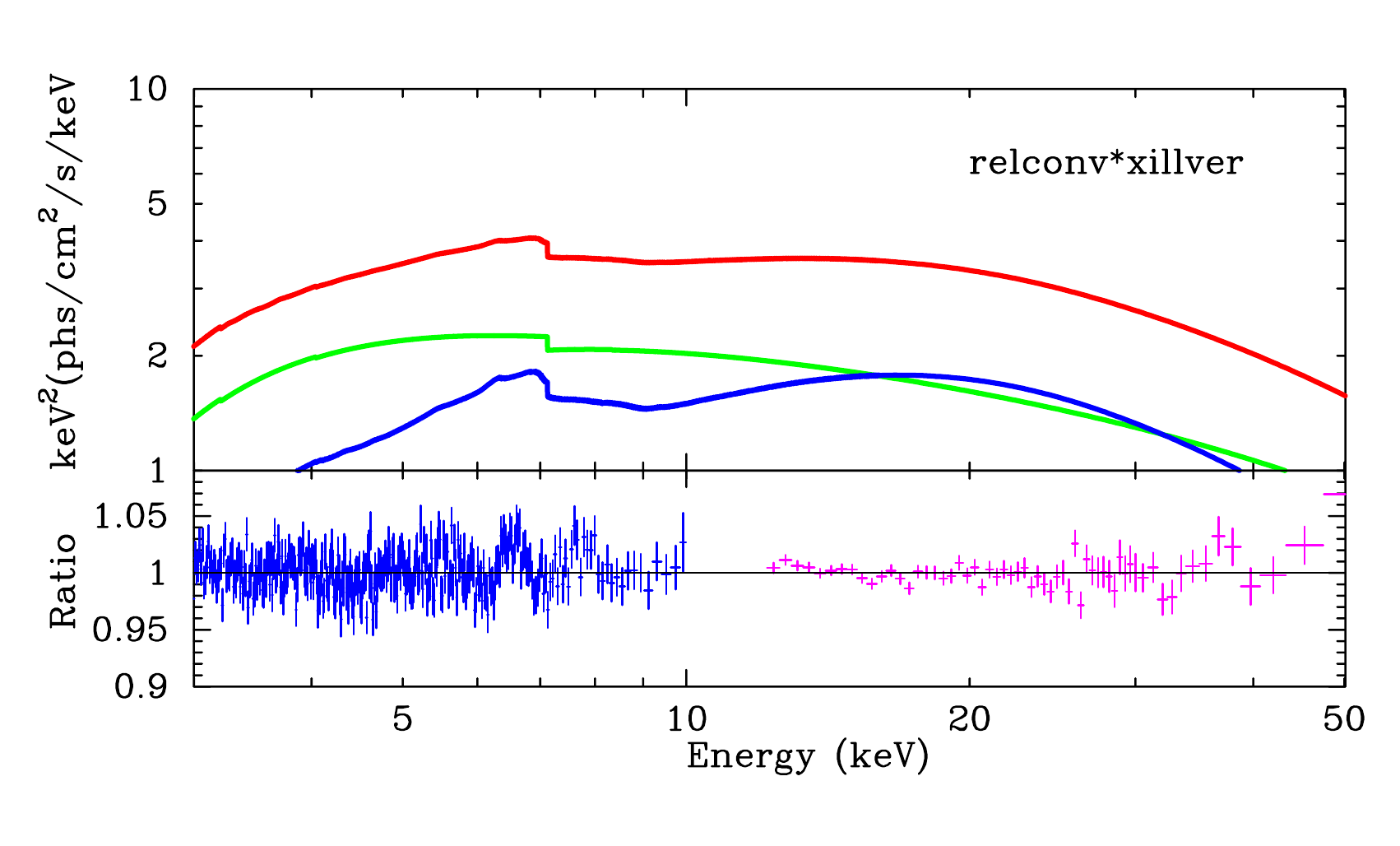}
\hspace{0.0cm}
\includegraphics[width=5.7cm,trim={1.7cm 0cm 2.6cm 17.5cm},clip]{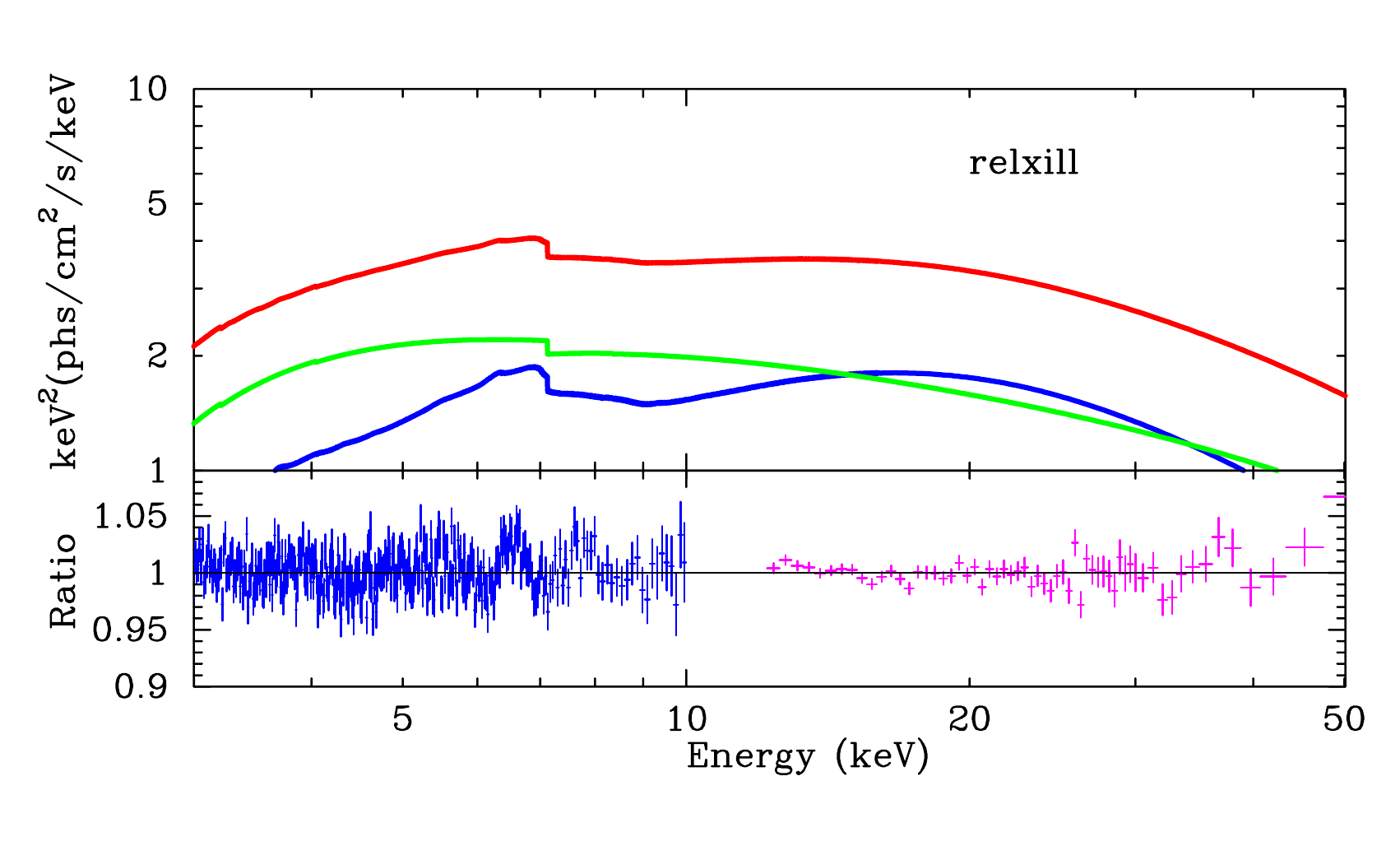}
\end{center}
\vspace{-0.4cm}
\caption{GRS~1915+105 -- Best-fit models (top quadrants) and data to best-fit model ratios (bottom quadrants) for the models {\sc relconv$\times$reflionx} (left panel), {\sc relconv$\times$xillver} (central panel), and {\sc relxill} (right panel). The red, blue, and greed curves correspond, respectively, to the total spectrum, the relativistic reflection component, and the power law component. \label{f-grs-mr}}
\vspace{0.2cm}
\begin{center}
\includegraphics[width=5.7cm,trim={0cm 0cm 1.0cm 0.2cm},clip]{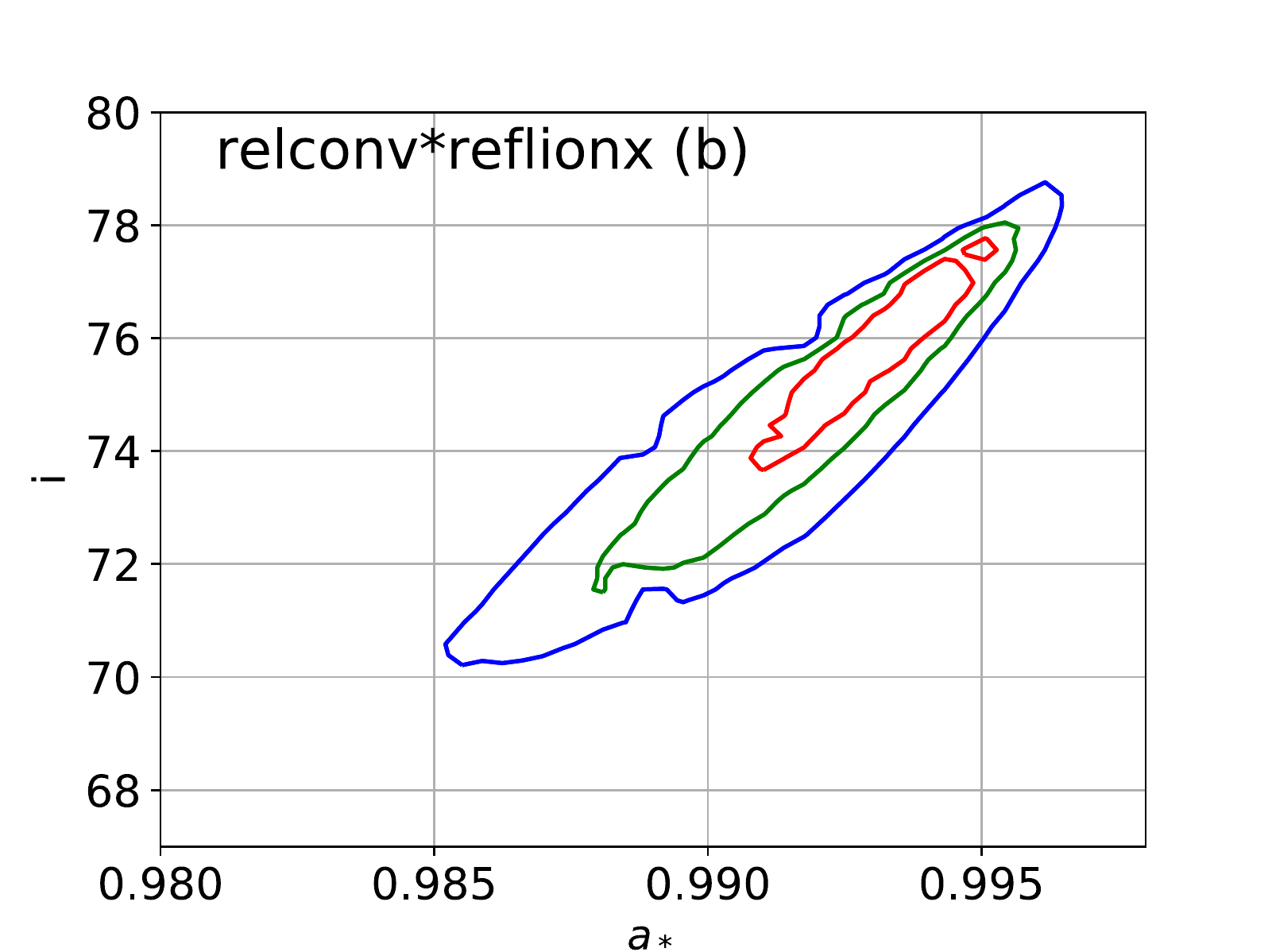}
\hspace{0.0cm}
\includegraphics[width=5.7cm,trim={0cm 0cm 1.0cm 0.2cm},clip]{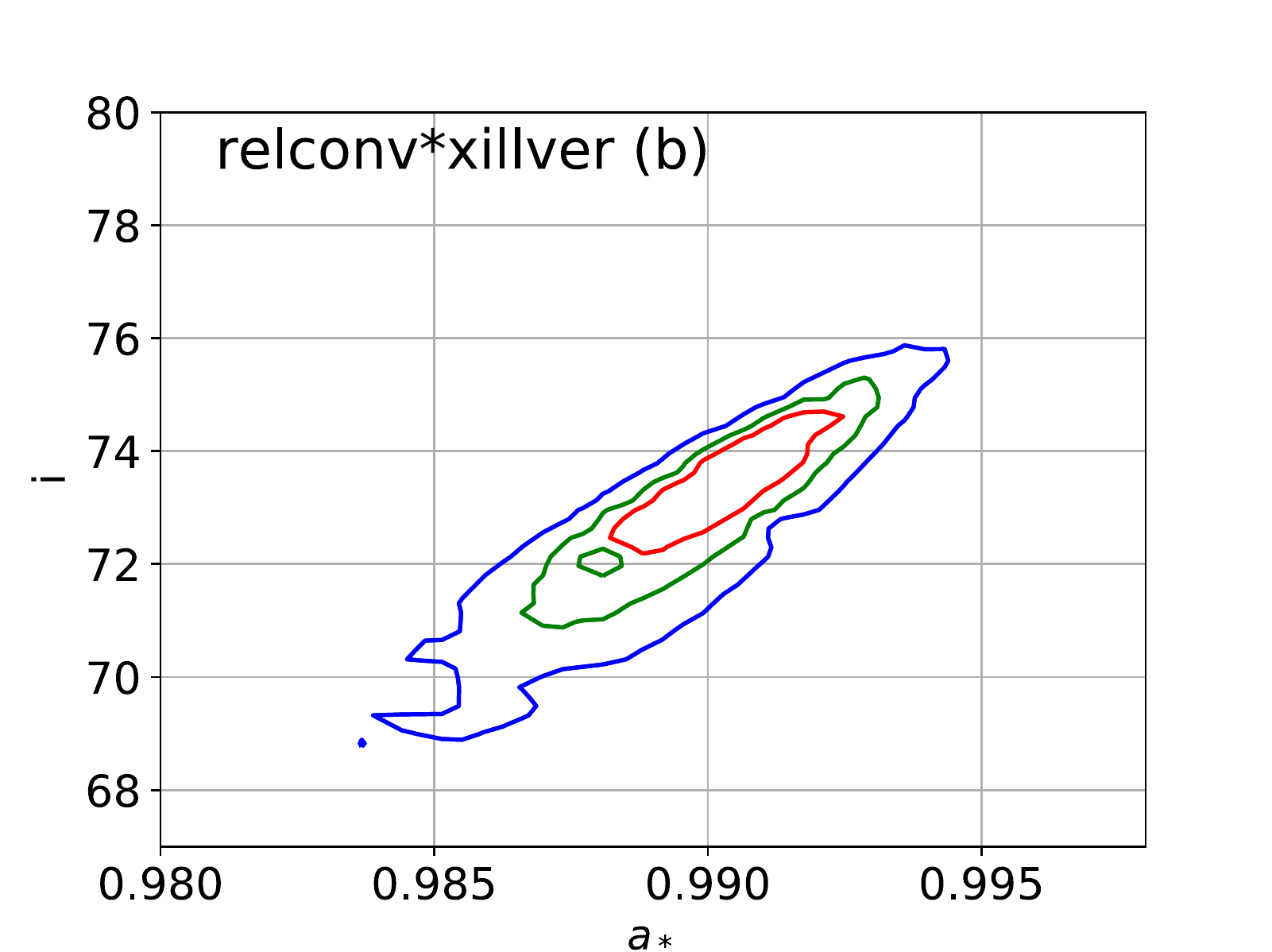}
\hspace{0.0cm}
\includegraphics[width=5.7cm,trim={0cm 0cm 1.0cm 0.2cm},clip]{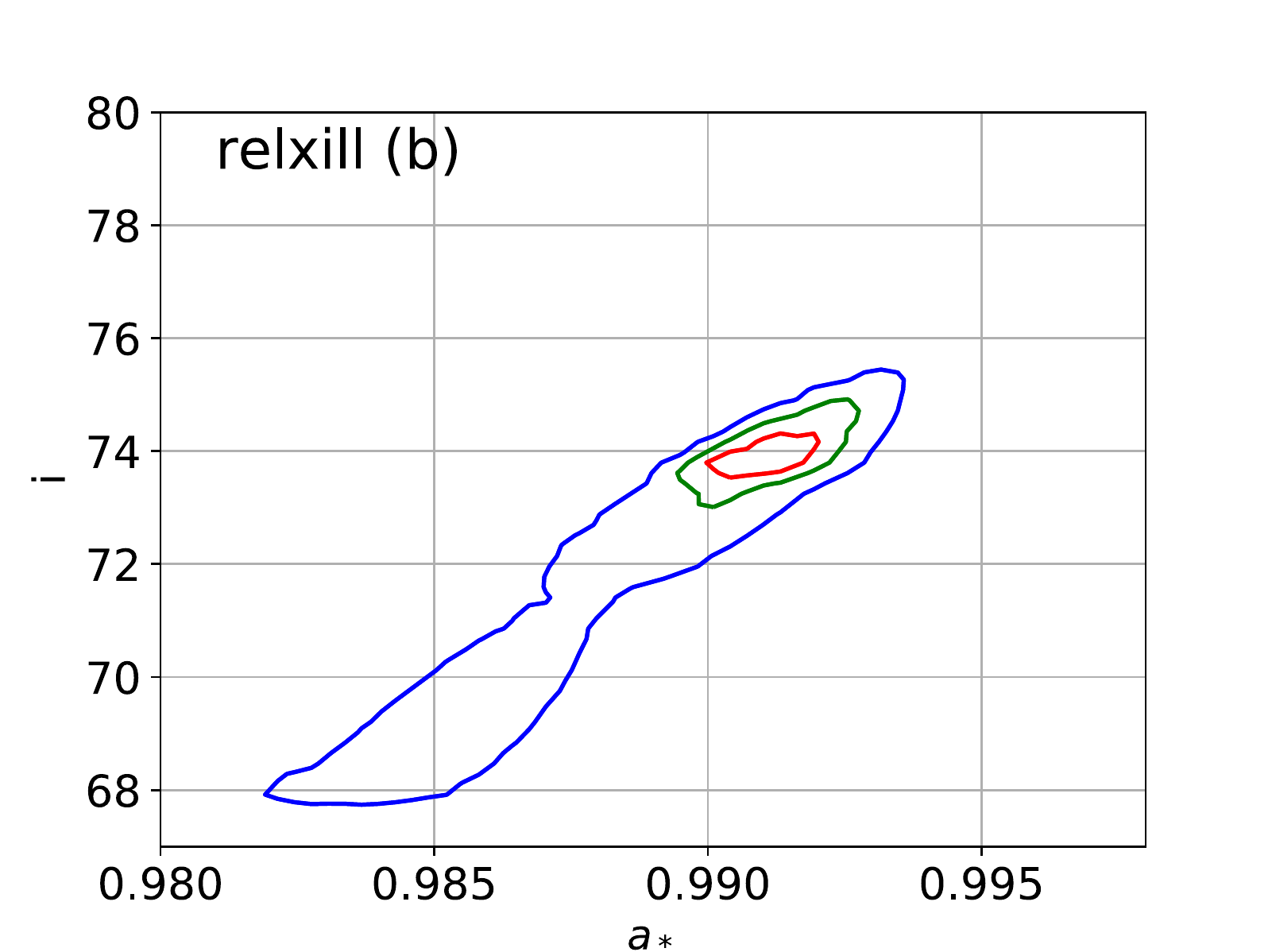}
\end{center}
\vspace{-0.2cm}
\caption{GRS~1915+105 -- Constraints on the black hole spin parameter $a_*$ and the inclination angle of the disk $i$ for the models {\sc relconv$\times$reflionx} (left panel), {\sc relconv$\times$xillver} (central panel), and {\sc relxill} (right panel). The red, greed, and blue curves correspond, respectively, to the 68\%, 90\%, and 99\% confidence level limits for two relevant parameters. \label{f-grs-ai}}
\vspace{0.2cm}
\begin{center}
\includegraphics[width=5.7cm,trim={0cm 0cm 1.0cm 0.2cm},clip]{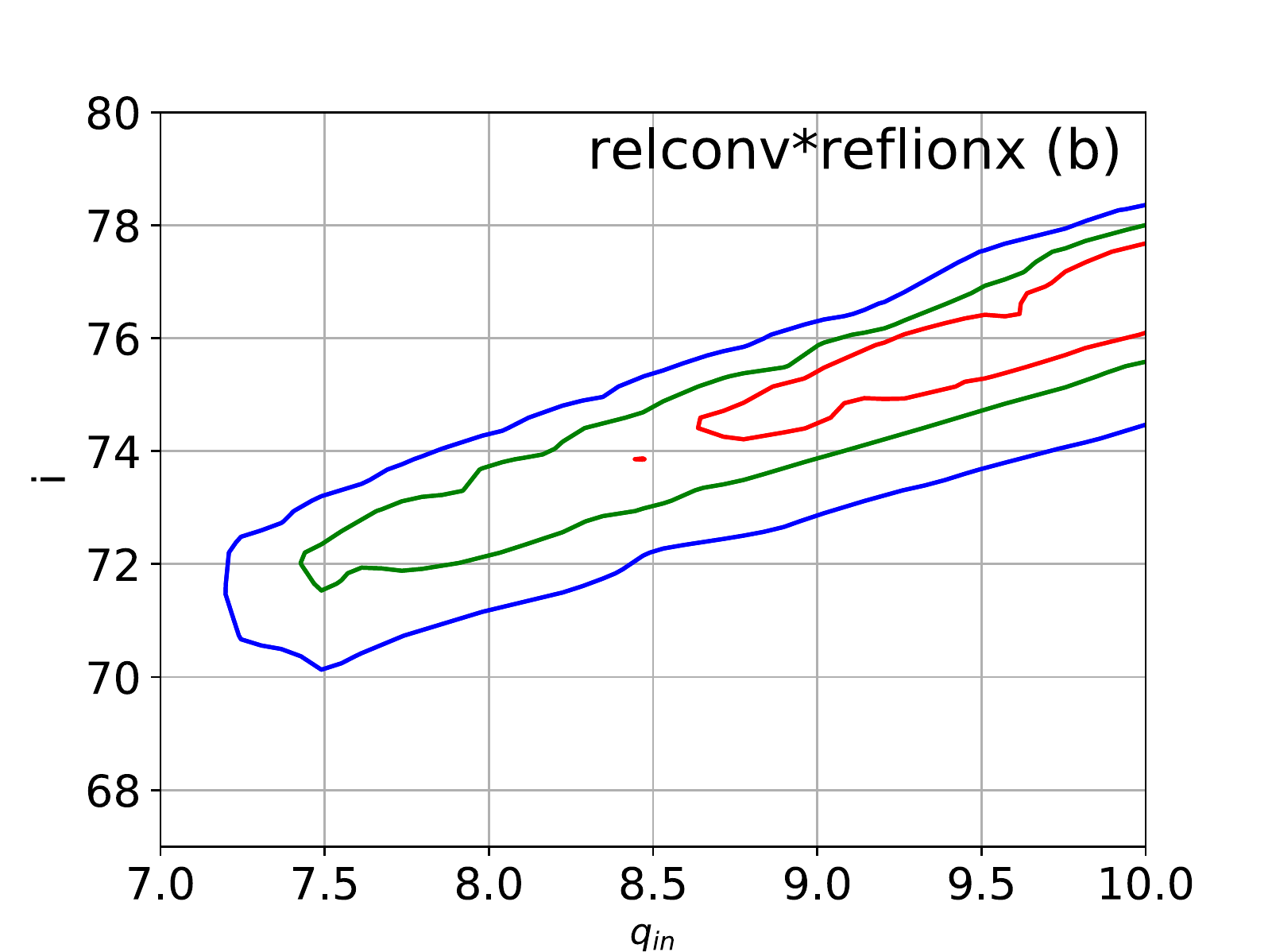}
\hspace{0.0cm}
\includegraphics[width=5.7cm,trim={0cm 0cm 1.0cm 0.2cm},clip]{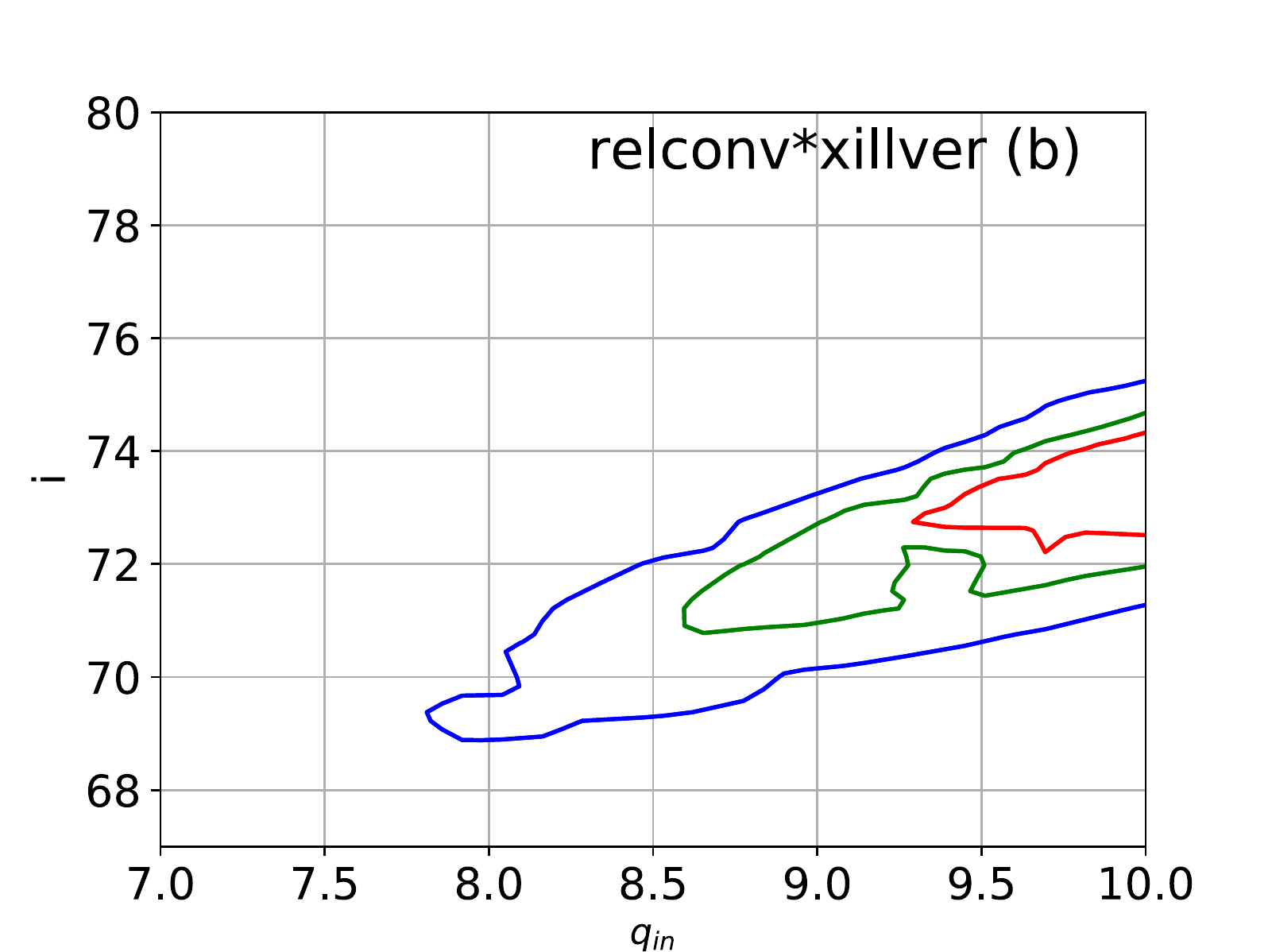}
\hspace{0.0cm}
\includegraphics[width=5.7cm,trim={0cm 0cm 1.0cm 0.2cm},clip]{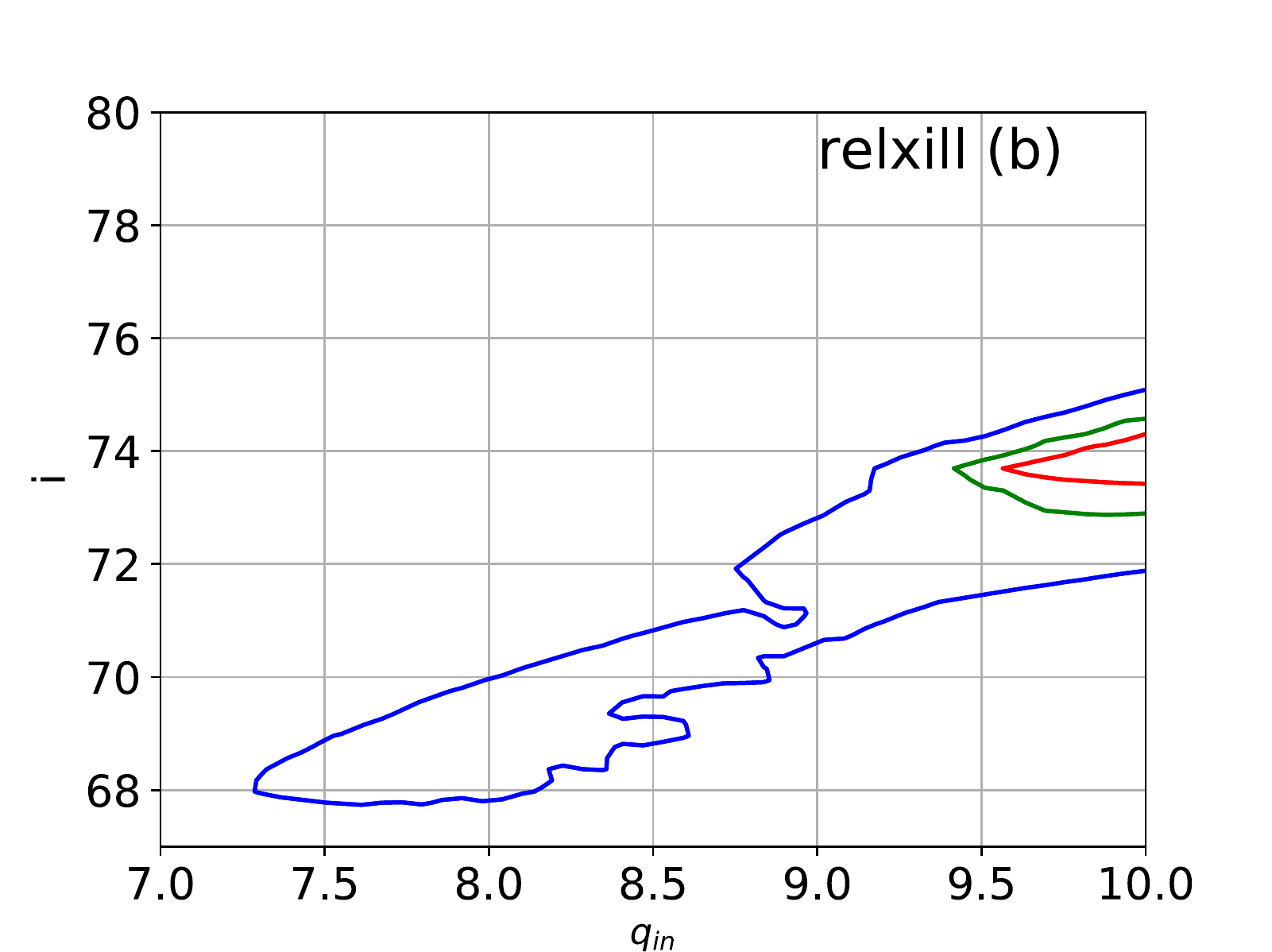}
\end{center}
\vspace{-0.2cm}
\caption{GRS~1915+105 -- Constraints on the inner emissivity index $q_{\rm in}$ and the inclination angle of the disk $i$ for the models {\sc relconv$\times$reflionx} (left panel), {\sc relconv$\times$xillver} (central panel), and {\sc relxill} (right panel). The red, greed, and blue curves correspond, respectively, to the 68\%, 90\%, and 99\% confidence level limits for two relevant parameters. \label{f-grs-qi}}
\end{figure*}

\subsection{Spectral analysis}

We use XSPEC v~12.10.1f~\citep{xspec}. The data can be fitted with an absorbed coronal and relativistic reflection components and no other components seem to be necessary~\citep{2009ApJ...706...60B,2019ApJ...884..147Z}. In XSPEC language, the model is

\vspace{0.15cm}
{\sc tbabs$\times$(cutoffpl + RR)} .
\vspace{0.15cm}

\noindent {\sc tbabs} models the Galactic absorption~\citep{2000ApJ...542..914W}, and its hydrogen column density $N_{\rm H}$ is left free in the fit. {\sc cutoffpl} describes a power-law component (photon index $\Gamma$) with a high energy exponential cut-off ($E_{\rm cut}$). We have three free parameters from this component in the fit: $\Gamma$, $E_{\rm cut}$, and the normalization of the component. If we fit the data with the model {\sc tbabs$\times$cutoffpl}, we find the data to best-fit model ratio shown in Fig.~\ref{f-grs-ratio}, where we can clearly see relativistic reflection features, with a broad iron line in the soft X-ray band and the Compton hump in the hard X-ray band. {\sc RR} indicates a relativistic reflection component and we employ three models: {\sc relconv$\times$reflionx}, {\sc relconv$\times$xillver}, and {\sc relxill}. We model the emissivity profile of the disk with a broken power-law and we have eight free parameters: black hole spin ($a_*$), inclination angle of the disk ($i$), inner emissivity index ($q_{\rm in}$), outer emissivity index ($q_{\rm out}$), breaking radius ($R_{\rm br}$), iron abundance ($A_{\rm Fe}$), ionization parameter ($\xi$), and normalization of the component. When we use {\sc relconv$\times$xillver} and {\sc relxill}, the reflection fraction is frozen to $-1$ because the coronal spectrum is described with {\sc cufoffpl}. The photon index and the high energy cut-off of the radiation illuminating the disk are tied to $\Gamma$ and $E_{\rm cut}$ in {\sc cufoffpl}.

Fig.~\ref{f-grs-mr} shows the best-fit models and the data to best-fit model ratios of the three scenarios. The best-fit values are reported in Tab.~\ref{t-grs}. Fig.~\ref{f-grs-ai} and Fig.~\ref{f-grs-qi} show, respectively, the constraints on the black hole spin $a_*$ and the inclination angle of the disk $i$ and on the inner emissivity index $q_{\rm in}$ and the inclination angle of the disk $i$. The red, greed, and blue curves correspond, respectively, to the 68\%, 90\%, and 99\% confidence level limits for two relevant parameters. A quick look already sees that the fit with {\sc relconv$\times$reflionx} is slightly worse, but all the measurements of the three models are perfectly consistent among them. A more detailed discussion on these fits is postponed to Section~\ref{s-dc}.


\begin{figure}
\begin{center}
\includegraphics[width=8.5cm,trim={1.0cm 1.0cm 3.7cm 10.0cm},clip]{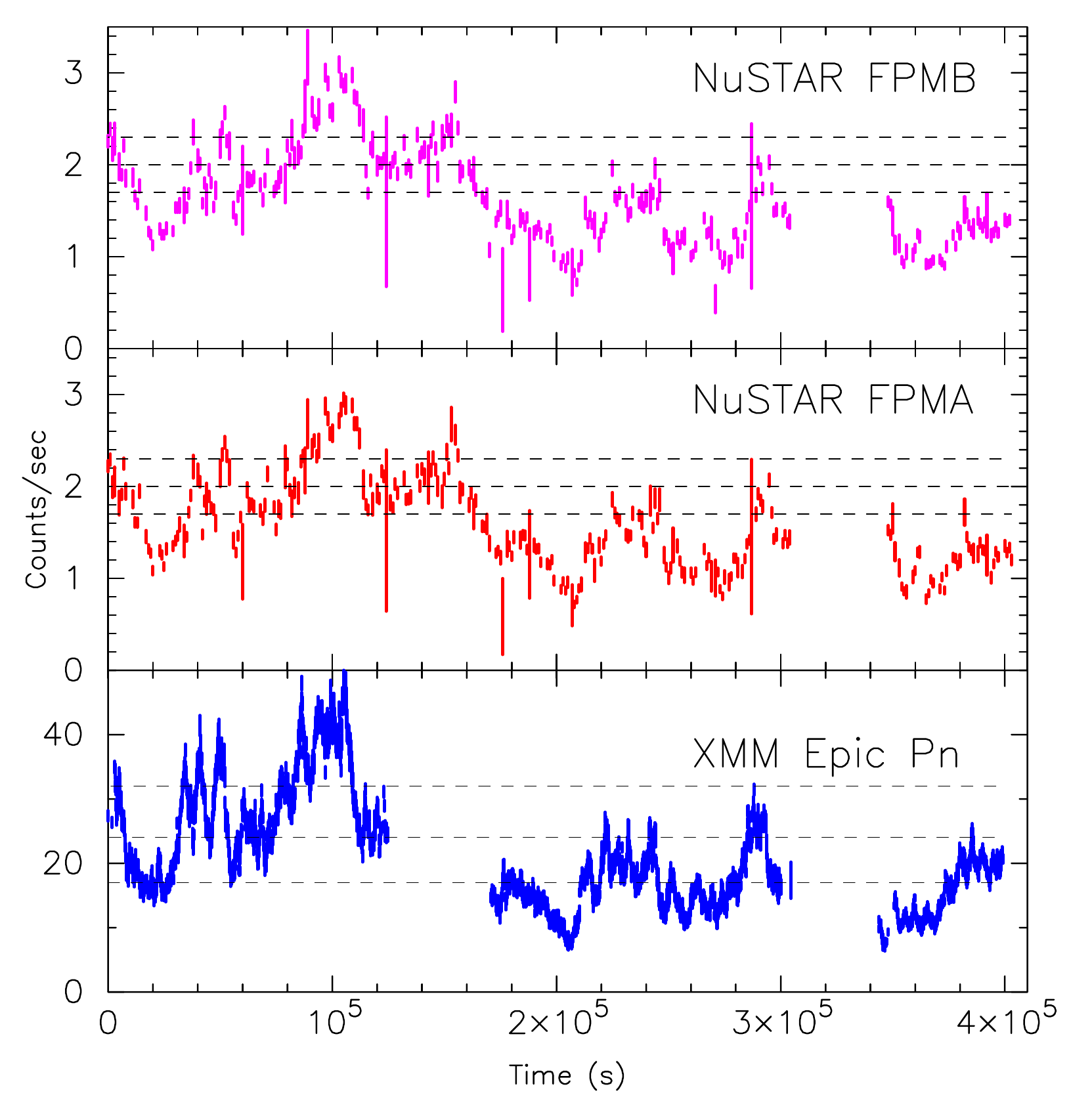}
\end{center}
\vspace{-0.2cm}
\caption{MCG--6--30--15 -- \textsl{NuSTAR}/FPMA, \textsl{NuSTAR}/FPMB, and \textsl{XMM-Newton}/EPIC-Pn light curves. The dashed horizontal lines separate the different flux states (low, medium, high, and very-high). \label{f-mcg-lc}}
\end{figure}

\begin{figure*}
\begin{center}
\includegraphics[width=8.5cm,trim={1.5cm 0cm 4.0cm 18.2cm},clip]{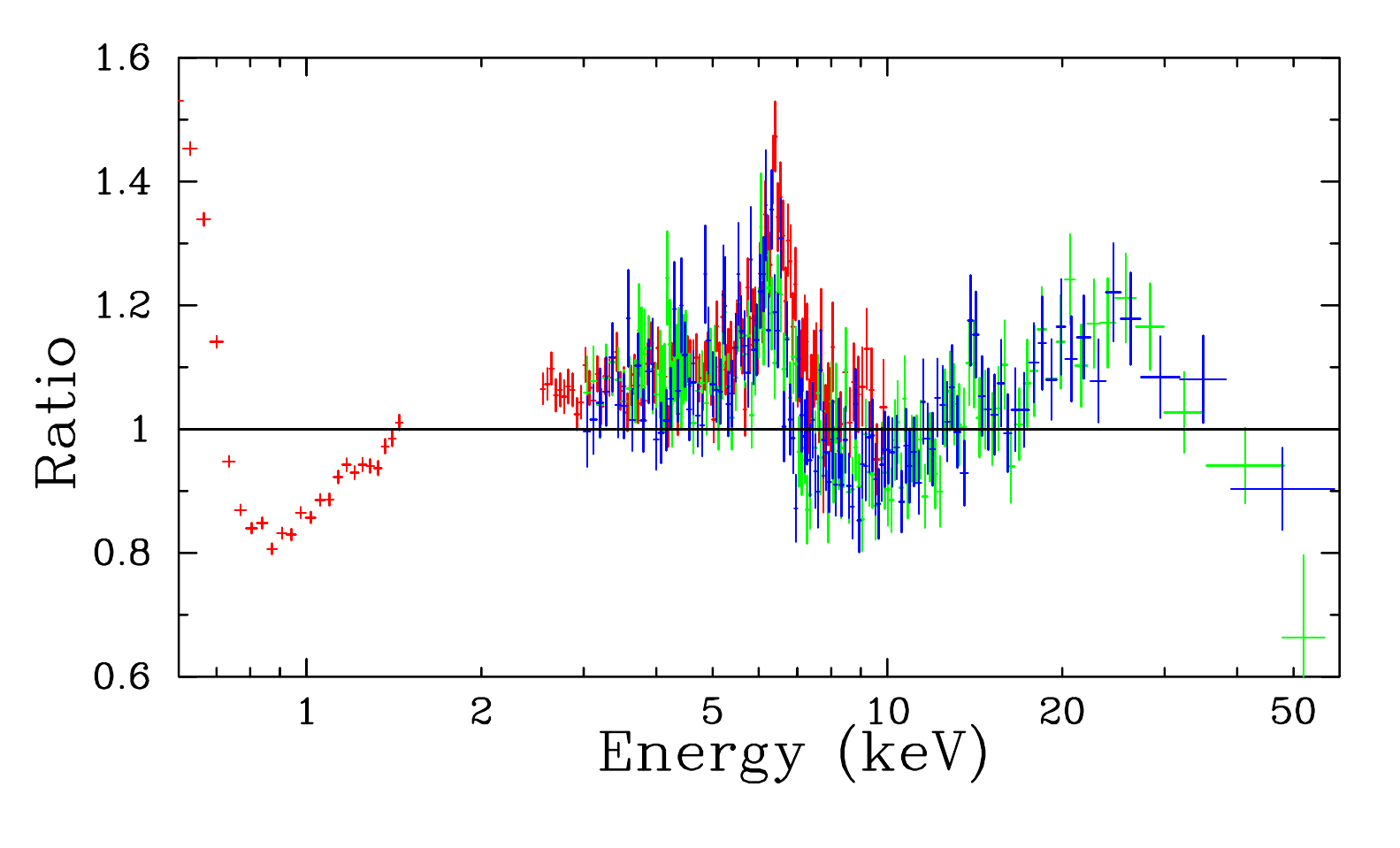}
\hspace{0.5cm}
\includegraphics[width=8.5cm,trim={1.5cm 0cm 4.0cm 18.2cm},clip]{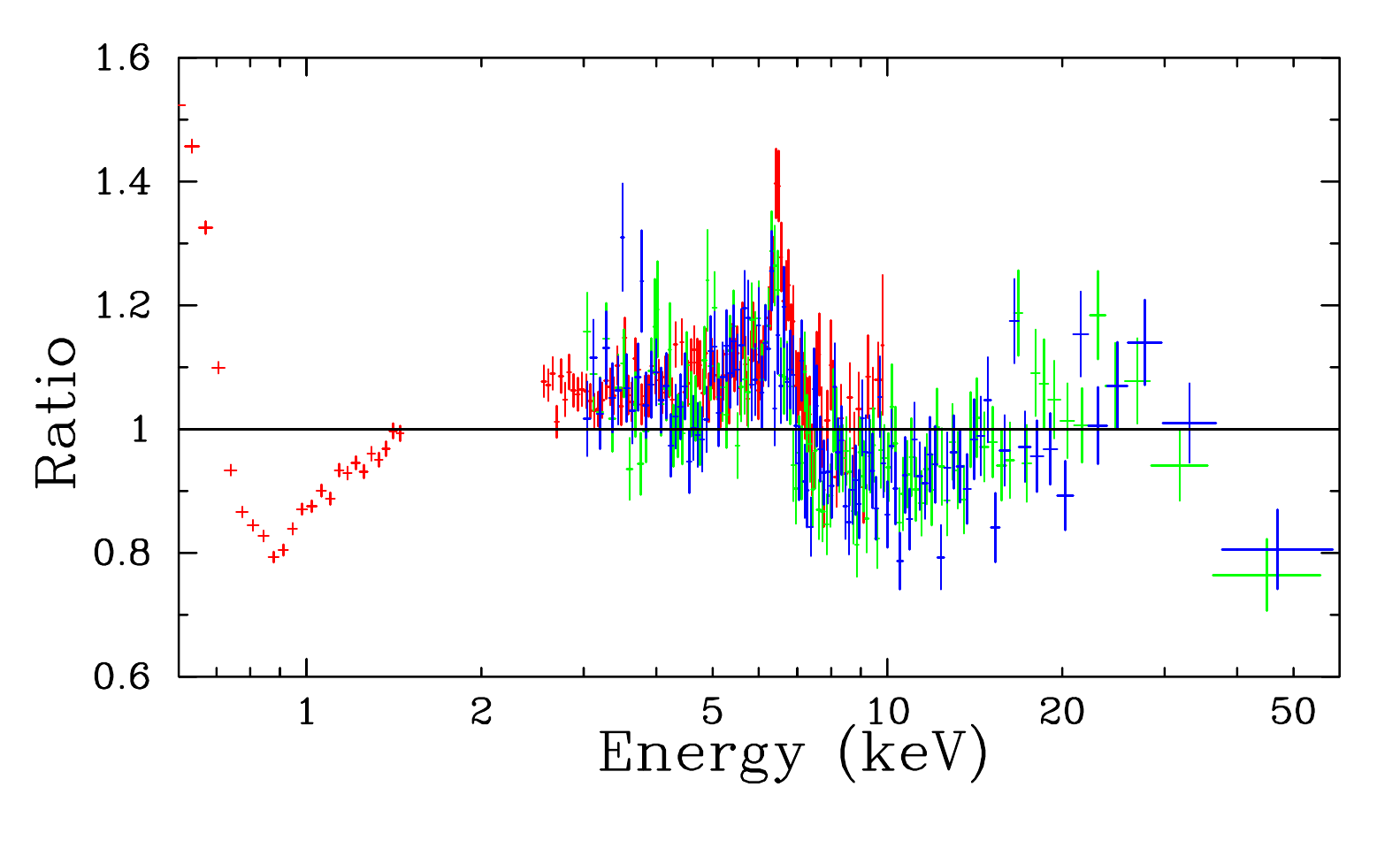} \\
\vspace{0.3cm}
\includegraphics[width=8.5cm,trim={1.5cm 0cm 4.0cm 18.2cm},clip]{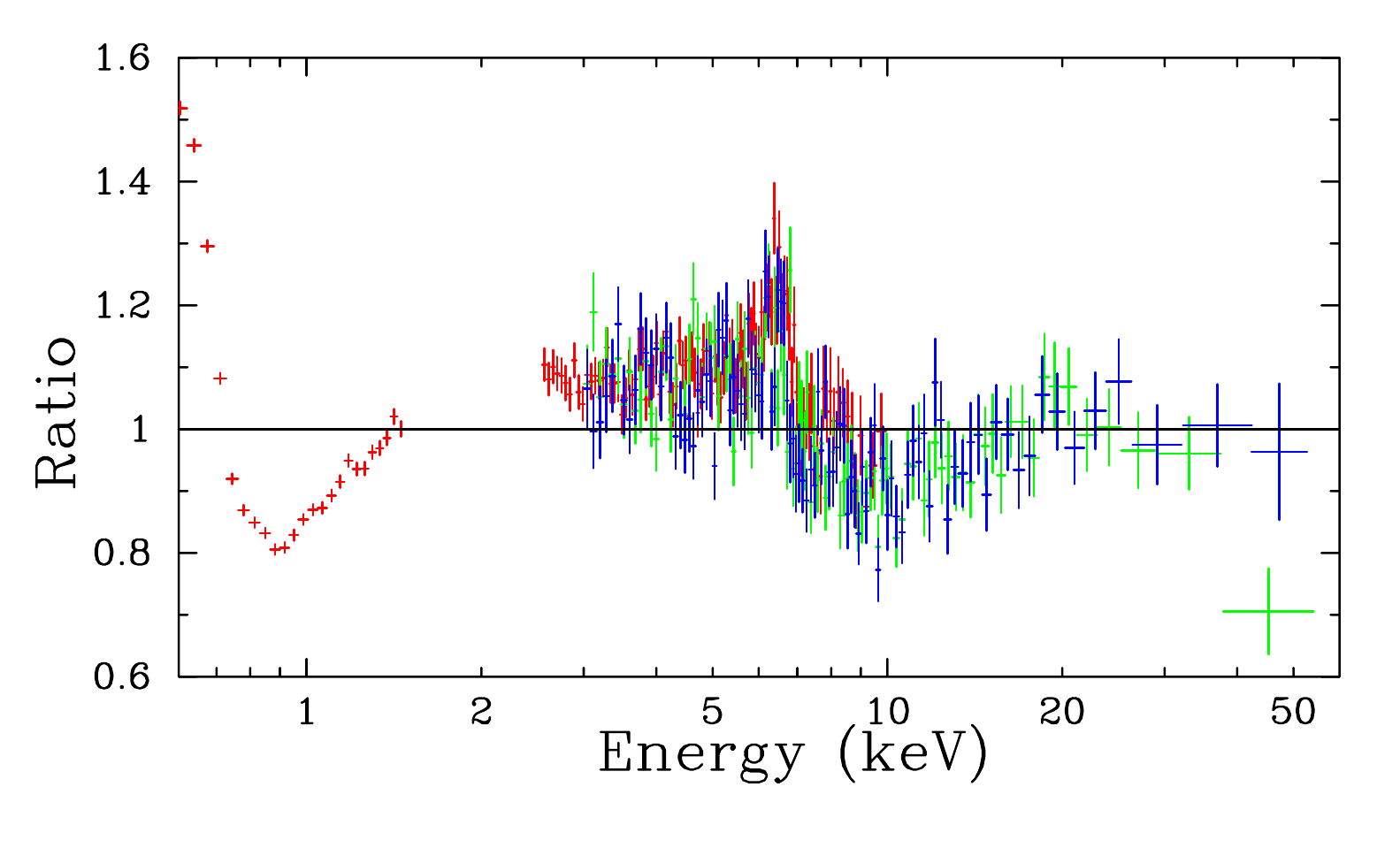}
\hspace{0.5cm}
\includegraphics[width=8.5cm,trim={1.5cm 0cm 4.0cm 18.2cm},clip]{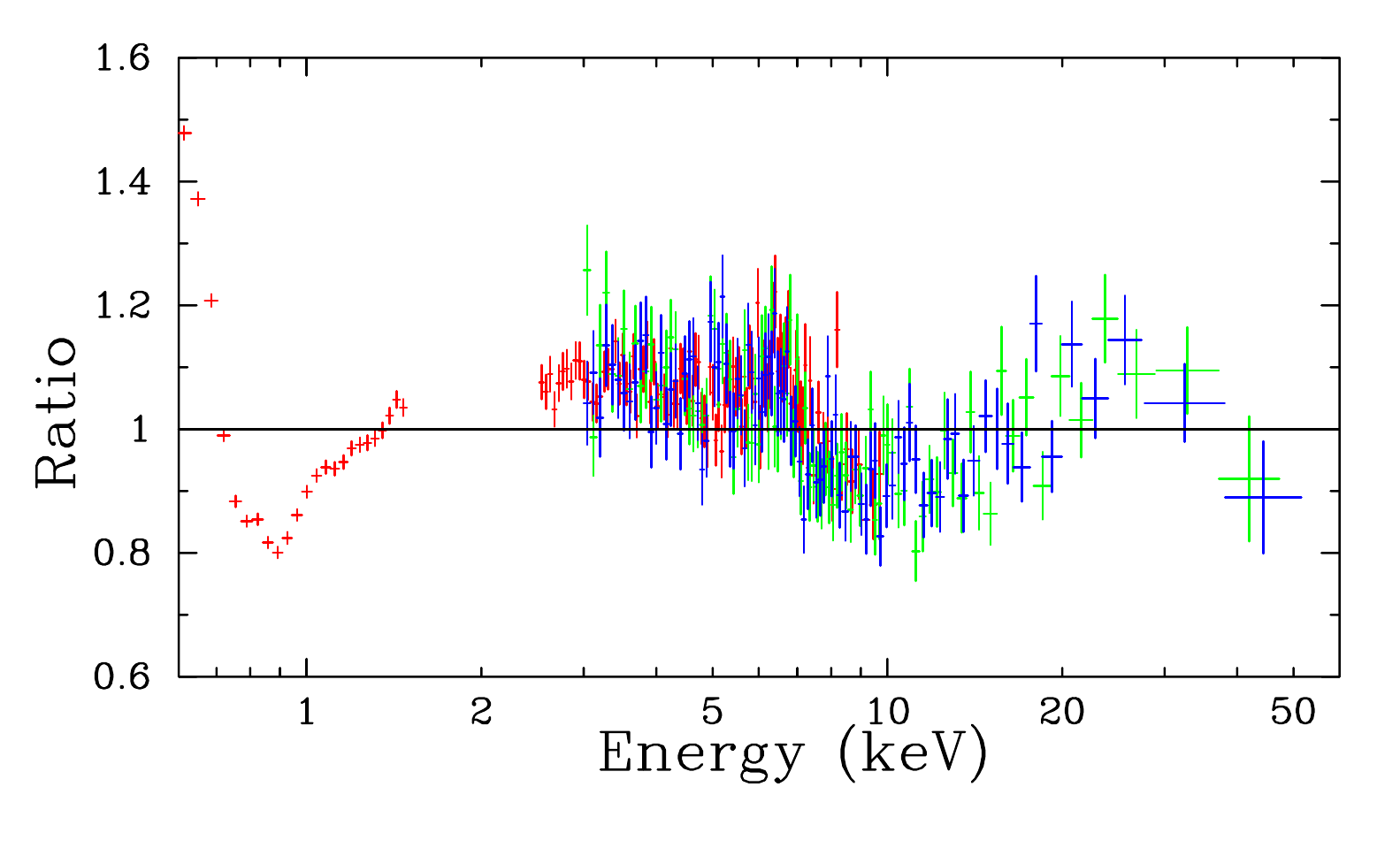}
\end{center}
\vspace{-0.4cm}
\caption{MCG--6--30--15 -- Data to best-fit model ratio for {\sc tbabs$\times$cutoffpl} for the low flux state (top left panel), medium flux state (top right panel), high flux state (bottom left panel), and very-high flux state (bottom right panel). Red crosses are used for \textsl{XMM-Newton} data, green crosses are used for \textsl{NuSTAR}/FPMA, and blue crosses are used for \textsl{NuSTAR}/FPMB. \label{f-mcg-ratio}}
\end{figure*}

\section{Reflection spectrum of MCG--6--30--15}\label{s-mcg}

MCG--6--30--15 is a very bright Seyfert~1 galaxy at redshift $z = 0.007749$. This source has been studied by many authors with different observations and different X-ray missions~\citep[see, e.g., ][]{1996MNRAS.282.1038I,1999A&A...341L..27G,2002MNRAS.335L...1F,2006ApJ...652.1028B,2007PASJ...59S.315M,2014ApJ...787...83M}. It is the source in which a relativistically broadened iron line was unambiguously detected for the first time~\citep{1995Natur.375..659T}. Indeed its spectrum often shows a very broad and prominent iron K$\alpha$ line. This makes MCG--6--30--15 an excellent candidate for a number of studies of relativistic reflection spectra. The source is very variable, so the data analysis requires some special attention. There are a few measurements in literature of the spin of the black hole in MCG--6--30--15 and they all agree in a very high value of the spin parameter $a_*$~\citep{2006ApJ...652.1028B,2014ApJ...787...83M,2019ApJ...875...56T}.

\subsection{Observations and data reduction}

\textsl{XMM-Newton} and \textsl{NuSTAR} observed simultaneously MCG--6--30--15 from 2013-1-29 for a total time of $\sim$315~ks and $\sim$360~ks, respectively (obs. ID~0693781201, 0693781301, and 0693781401 for \textsl{XMM-Newton} and obs. ID~60001047002, 60001047003, and 60001047005 for \textsl{NuSTAR}); see Tab.~\ref{t-obsid}. The first study of these data was reported in \citet{2014ApJ...787...83M}, where the estimates of the spin parameter and the inclination angle of the disk are, respectively, $a_* = 0.91_{-0.07}^{+0.06}$ and $i = 33^\circ \pm 3^\circ$.

For the data reduction, we follow \citet{2019ApJ...875...56T}. Here we only outline the main passages. For the \textsl{XMM-Newton} data, we only use Pn data because they are of higher quality, while the MOS data are significantly affected by pile up. We use SAS version~16.0.0 to convert raw data into event files. The latter are combined into a single FITS file using the ftool \texttt{fmerge}; good time intervals (GTIs) are generated using \texttt{tabtigen} and then used to filter the event files. The source region is a circle of radius 40~arcsec and the background region is a circle of radius 50~arcsec. Spectra are rebinned to oversample the instrumental resolution by at least a factor of 3 and have at least 50~counts in each background subtracted bin. As for the \textsl{NuSTAR} data, we use the HEASOFT task \texttt{nupipeline} with CALDB version 20180312 to generate the event files. The source region is a circle of radius 70~arcsec and the background region is a circle of radius 100~arcsec. Spectra are rebinned to guarantee at least 70~counts per bin.

MCG--6--30--15 is very variable and it is thus necessary to use simultaneous data in order to properly take the variability into account. Here we use the ftool \texttt{mgtime} to find the common GTIs of the two telescopes.

\subsection{Spectral analysis}

Since the source is very variable during the \textsl{XMM-Newton} and \textsl{NuSTAR} observations, see the light curves in Fig.~\ref{f-mcg-lc}, we proceed as in \citet{2019ApJ...875...56T}. We arrange the data into four groups according to the flux state of the source (low, medium, high, and very-high flux states) requiring similar spectral data count for every flux state. We use XSPEC v~12.10.1f~\citep{xspec}.

The \textsl{XMM-Newton} data show a spurious emission around 2~keV~\citep{2014ApJ...787...83M}. This is interpreted as an effect of the golden edge in the response file due to miscalibration in the long-term charge transfer inefficiency. We cannot fit this feature with an ad hoc Gaussian because, otherwise, this would modify the way in which warm absorbers and ionized reflectors reproduce the data. We thus ignore the 1.5-2.5~keV energy band in the \textsl{XMM-Newton} data.

The search for the model is described in \citet{2019ApJ...875...56T}. If we fit the data with an absorbed power law, we clearly see typical relativistic reflection features: a broad iron line peaked around 6~keV, a soft excess below 1~keV, and a Compton hump peaked at 20-30~keV, see Fig.~\ref{f-mcg-ratio}.

The final model is

\vspace{0.15cm}
{\sc tbabs$\times$warmabs$_1$$\times$warmabs$_2$$\times$dustyabs} 

{\sc \hspace{0.4cm} $\times$(cutoffpl + RR + NR + zgauss + zgauss)} .
\vspace{0.15cm}

\noindent {\sc tbabs} describes the Galactic absorption and we fix the column density $N_{\rm H} = 3.9 \cdot 10^{20}$~cm$^{-2}$~\citep{1990ARA&A..28..215D}. {\sc warmabs$_1$} and {\sc warmabs$_2$} describe two ionized absorbers; their tables are generated with {\sc xstar}~v2.41. {\sc dustyabs} describes a neutral absorber and only modifies the soft X-ray band~\citep{2001ApJ...554L..13L}. {\sc cutoffpl} describes the direct spectrum from the corona. {\sc RR} and {\sc NR} indicate, respectively, a relativistic and a non-relativistic reflection component. When we use {\sc reflionx}, we have {\sc RR=relconv$\times$reflionx} and {\sc NR=reflionx}. When {\sc RR} is either {\sc relconv$\times$xillver} or {\sc relxill}, {\sc NR} is {\sc xillver}. The non-relativistic reflection component is interpreted as some cold material faraway from the black hole, so we set its ionization parameter to the minimum value allowed by the model. We model the emissivity profile of the accretion disk either with a simple power-law [indicated with (s) in our tables and figures] and a broken power-law [indicated with (b) in our tables and figures]. {\sc zgauss} is used to describe a narrow emission line around 0.8~keV, which is interpreted as an oxygen line, and a narrow absorption line at 1.2~keV, interpreted as blueshifted oxygen absorption caused by the presence of relativistic outflows~\citep{1997ApJ...489L..25L}.

Note that some model parameters cannot vary on a time scale of a few days and therefore their value should be the same over different flux states. These are the black hole spin $a_*$, the inclination angle of the disk $i$, and the iron abundance $A_{\rm Fe}$ as well as all the model parameters related to material at larger radii and appearing in {\sc dustyabs}, {\sc NR}, and {\sc zgauss}.

The best-fit values are reported in Tab.~\ref{t-mcg1} ({\sc relconv$\times$reflionx}), Tab.~\ref{t-mcg2} ({\sc relconv$\times$xillver}), and Tab.~\ref{t-mcg3} ({\sc relxill}). In every table, we show the fit with the emissivity profile of the disk modeled by a simple power-law (left) and by a broken power-law (right). For every model, there are four columns, corresponding to the low, medium, high, and very-high flux states, respectively. Fig.~\ref{f-grs-mr} shows the best-fit models of the low flux states and the data to best-fit model ratios. Fig.~\ref{f-grs-ai} is for the constraints on the spin parameter $a_*$ and the inclination angle of the disk $i$ for the six scenarios considered in our analysis. Unlike the analysis of the \textsl{Suzaku} observation of GRS~1915+105, here we see some differences and the best-fit values from different models are not always consistent.


\begin{table*}
\centering
{\renewcommand{\arraystretch}{1.25}
\begin{tabular}{l|cccc|cccc}
\hline\hline
Model & \multicolumn{4}{c}{{\sc relconv$\times$reflionx} (s)} & \multicolumn{4}{c}{{\sc relconv$\times$reflionx} (b)} \\
\hline
Flux state & 1 & 2 & 3 & 4 & 1 & 2 & 3 & 4 \\
\hline
{\sc tbabs} &&&& &&&& \\
$N_{\rm H} / 10^{22}$ cm$^{-2}$ & \multicolumn{4}{c}{$0.039^\star$} & \multicolumn{4}{c}{$0.039^\star$} \\
\hline
{\sc warmabs$_1$} &&&& \\
$N_{\rm H \, 1} / 10^{22}$ cm$^{-2}$ & $0.323^{+0.015}_{-0.014}$ & $0.426^{+0.023}_{-0.022}$ & $0.388^{+0.021}_{-0.020}$ & $0.09^{+0.05}_{-0}$ 
& $0.33_{-0.06}^{+0.05}$ & $0.44_{-0.07}^{+0.06}$ & $0.39_{-0.07}^{+0.09}$ & $0.38_{-0.05}^{+0.09}$ \\
$\log\xi_1$ & $2.258^{+0.029}_{-0.023}$ & $2.391^{+0.024}_{-0.024}$ & $2.397^{+0.026}_{-0.025}$ & $1.38^{+0.22}_{-0.30}$ 
& $2.23_{-0.08}^{+0.10}$ & $2.33_{-0.07}^{+0.07}$ & $2.35_{-0.10}^{+0.10}$ & $2.45_{-0.07}^{+0.11}$ \\
\hline
{\sc warmabs$_2$} &&&& \\
$N_{\rm H \, 2} / 10^{22}$ cm$^{-2}$ & $2.58^{+0.16}_{-0.15}$ & $1.29^{+0.11}_{-0.13}$ & $1.45^{+0.11}_{-0.10}$ & $0.386^{+0.036}_{-0.021}$ 
& $1.9_{-0.4}^{+0.5}$ & $1.7_{-0.4}^{+0.9}$ & $1.9_{-0.4}^{+0.4}$ & $1.8_{-0.9}^{+1.1}$ \\
$\log\xi_2$ & $3.347^{+0.012}_{-0.012}$ & $3.324^{+0.020}_{-0.021}$ & $3.291^{+0.020}_{-0.019}$ & $2.40^{+0.03}_{-0.03}$ 
& $3.28_{-0.05}^{+0.05}$ & $3.32_{-0.05}^{+0.06}$ & $3.29_{-0.05}^{+0.05}$ & $3.41_{-0.10}^{+0.06}$ \\
\hline
{\sc dustyabs} &&&& \\
$\log \big( N_{\rm Fe} / 10^{21}$ cm$^{-2} \big)$ & \multicolumn{4}{c}{$17.6632^{+0.0030}_{-0.0033}$} 
& \multicolumn{4}{c}{$17.628_{-0.017}^{+0.019}$} \\
\hline
{\sc cutoffpl} &&&& \\
$\Gamma$ & $1.8571^{+0.0025}_{-0.0025}$ & $1.8711^{+0.0025}_{-0.0025}$ & $1.9269^{+0.0026}_{-0.0026}$ & $1.961^{+0.003}_{-0.003}$ 
& $1.843_{-0.013}^{+0.012}$ & $1.868_{-0.011}^{+0.012}$ & $1.924_{-0.012}^{+0.012}$ & $1.948_{-0.014}^{+0.013}$ \\
$E_{\rm cut}$ [keV] & $207^{+26}_{-19}$ & $96^{+6}_{-5}$ & $90^{+5}_{-4}$ & $90^{+5}_{-4}$
& 
$218_{-56}^{+74}$ & $114_{-20}^{+32}$ & $108_{-19}^{+28}$ & $147_{-34}^{+63}$ \\
$N_\text{\sc cutoffpl}$~$(10^{-3})$ & $8.42^{+0.13}_{-0.13}$ & $12.47^{+0.15}_{-0.15}$ & $18.03^{+0.20}_{-0.20}$ & $26.93^{+0.20}_{-0.09}$ 
& $7.8_{-0.4}^{+0.6}$ & $11.8_{-0.5}^{+0.5}$ & $16.7_{-0.8}^{+0.8}$ & $23.2_{-1.1}^{+1.2}$ \\ 
\hline
{\sc relconv} &&&& \\
$q_{\rm in}$ & $3.03^{+0.04}_{-0.20}$ & $2.71^{+0.05}_{-0.05}$ & $2.79^{+0.05}_{-0.05}$ & $2.5^{+0.3}_{-0.3}$
& $5_{\rm -(P)}^{\rm +(P)}$ & $3.6_{-0.6}^{+0.6}$ & $4.4_{-1.1}^{+5.2}$ & $6.0_{-1.2}^{+2.3}$ \\
$q_{\rm out}$ & $= q_{\rm in}$ & $= q_{\rm in}$ & $= q_{\rm in}$ & $= q_{\rm in}$ 
& $2.65_{-0.23}^{+0.22}$ & $1.9_{-0.4}^{+0.7}$ & $2.3_{-0.8}^{+0.4}$ & $1.8_{-0.7}^{+0.6}$ \\
$R_{\rm br}$ [$M$] & -- & -- & -- & --
& $4.3_{-2.2}^{+2.2}$ & $20_{-13}$ & $9_{-4}^{\rm +(P)}$ & $9_{-4}^{\rm +(P)}$ \\
$i$ [deg] & \multicolumn{4}{c}{$24.2^{+1.6}_{-1.7}$} 
& \multicolumn{4}{c}{$19_{-5}^{+5}$} \\
$a_*$ & \multicolumn{4}{c}{$0.83^{+0.03}_{-0.04}$} 
& \multicolumn{4}{c}{$0.88_{-0.08}^{+0.04}$} \\
\hline
{\sc reflionx} &&&& \\
$z$ & \multicolumn{4}{c}{$0.007749^\star$} & \multicolumn{4}{c}{$0.007749^\star$} \\
$\log\xi$ & $2.744^{+0.012}_{-0.003}$ & $2.54^{+0.06}_{-0.03}$ & $2.505^{+0.051}_{-0.025}$ & $2.094^{+0.047}_{-0.017}$ 
& $2.78_{-0.04}^{+0.04}$ & $2.73_{-0.03}^{+0.03}$ & $2.74_{-0.03}^{+0.04}$ & $2.80_{-0.05}^{+0.06}$ \\
$A_{\rm Fe}$ & \multicolumn{4}{c}{$3.17^{+0.08}_{-0.07}$} 
& \multicolumn{4}{c}{$2.9_{-0.4}^{+0.4}$} \\
$N_\text{\sc reflionx}$~$(10^{-6})$ & $0.111^{+0.059}_{-0.013}$ & $0.236^{+0.033}_{-0.019}$ & $0.370^{+0.043}_{-0.025}$ & $1.38^{+0.03}_{-0.09}$ 
& $0.133_{-0.015}^{+0.016}$ & $0.195_{-0.018}^{+0.021}$ & $0.30_{-0.05}^{+0.04}$ & $0.33_{-0.04}^{+0.04}$ \\ 
\hline
{\sc reflionx}$'$ &&&& \\
$\log\xi'$ & \multicolumn{4}{c}{$0^\star$} & \multicolumn{4}{c}{$0^\star$} \\
$N'_\text{\sc reflionx}$~$(10^{-6})$ & \multicolumn{4}{c}{$69^{+4}_{-4}$} 
& \multicolumn{4}{c}{$59_{-8}^{+8}$} \\
\hline
{\sc zgauss} &&&& &&&&\\
$E_{\rm line}$ [keV] & \multicolumn{4}{c}{$1.261^{+0.006}_{-0.012}$} 
& \multicolumn{4}{c}{$1.260_{-0.010}^{+0.011}$} \\
\hline
$\chi^2$/dof & \multicolumn{4}{c}{$3171.36/2691 = 1.17850$} & \multicolumn{4}{c}{$3130.26/2683 = 1.16670$} \\
\hline\hline
\end{tabular}
\vspace{0.2cm}
\caption{MCG--6--30--15 -- Summary of the best-fit values for the models {\sc relconv$\times$reflionx}~(s) and {\sc relconv$\times$reflionx}~(b). The reported uncertainties correspond to the 90\% confidence level for one relevant parameter ($\Delta\chi^2 = 2.71$). $\xi$, $\xi'$, $\xi_1$, and $\xi_2$ in units of erg~cm~s$^{-1}$. In {\sc reflionx}, the fitting parameter is $\xi$, not $\log\xi$ as in {\sc xillver} and {\sc relxill}, but it is converted into $\log\xi$ in the table in order to facilitate the comparison with the other models. See the text for the details. \label{t-mcg1}}
}
\end{table*}

\begin{table*}
\centering
{\renewcommand{\arraystretch}{1.25}
\begin{tabular}{l|cccc|cccc}
\hline\hline
Model & \multicolumn{4}{c}{{\sc relconv$\times$xillver} (s)} & \multicolumn{4}{c}{{\sc relconv$\times$xillver} (b)} \\
\hline
Flux state & 1 & 2 & 3 & 4 & 1 & 2 & 3 & 4 \\
\hline
{\sc tbabs} &&&& &&&& \\
$N_{\rm H} / 10^{22}$ cm$^{-2}$ & \multicolumn{4}{c}{$0.039^\star$} & \multicolumn{4}{c}{$0.039^\star$} \\
\hline
{\sc warmabs$_1$} &&&& \\
$N_{\rm H \, 1} / 10^{22}$ cm$^{-2}$ & $0.25_{-0.04}^{+0.07}$ & $1.18_{-0.07}^{+0.09}$ & $1.36_{-0.06}^{+0.07}$ & $1.10_{-0.09}^{+0.07}$ 
& $0.669_{-0.030}^{+0.012}$ & $0.10_{\rm -(P)}^{+0.19}$ & $1.03_{-0.06}^{+0.06}$ & $0.64_{-0.07}^{+0.05}$ \\
$\log\xi_1$ & $1.91_{-0.09}^{+0.18}$ & $2.005_{-0.021}^{+0.023}$ & $2.01_{-0.021}^{+0.022}$ & $1.982_{-0.033}^{+0.025}$ 
& $1.843_{-0.010}^{+0.043}$ & $1.84_{-0.25}^{+0.53}$ & $1.914_{-0.019}^{+0.031}$ & $1.77_{-0.05}^{+0.04}$ \\
\hline
{\sc warmabs$_2$} &&&& \\
$N_{\rm H \, 2} / 10^{22}$ cm$^{-2}$ & $1.00_{-0.10}^{+0.11}$ & $0.060_{-0.004}^{+0.008}$ & $0.23_{-0.16}^{+0.43}$ & $0.06_{\rm -(P)}^{+0.10}$ 
& $0.487_{-0.011}^{+0.031}$ & $1.07_{-0.07}^{+0.08}$ & $0.59_{-0.19}^{+0.23}$ & $0.32_{-0.06}^{+0.08}$ \\
$\log\xi_2$ & $2.00_{-0.05}^{+0.05}$ & $2.14_{-0.38}^{+0.08}$ & $3.20_{-0.19}^{+0.27}$ & $2.7_{-0.4}^{\rm +(P)}$ 
& $1.86_{-0.03}^{+0.06}$ & $1.900_{-0.024}^{+0.037}$ & $3.25_{-0.08}^{+0.06}$ & $2.51_{-0.14}^{+0.11}$ \\
\hline
{\sc dustyabs} &&&& \\
$\log \big( N_{\rm Fe} / 10^{21}$ cm$^{-2} \big)$ & \multicolumn{4}{c}{$17.391_{-0.022}^{+0.030}$} 
& \multicolumn{4}{c}{$17.41_{-0.04}^{+0.03}$} \\
\hline
{\sc cutoffpl} &&&& \\
$\Gamma$ & $1.949_{-0.016}^{+0.009}$ & $1.973_{-0.007}^{+0.013}$ & $2.045_{-0.012}^{+0.009}$ & $2.056_{-0.009}^{+0.007}$ 
& $1.968_{-0.015}^{+0.014}$ & $1.987_{-0.013}^{+0.013}$ & $2.025_{-0.015}^{+0.013}$ & $2.029_{-0.014}^{+0.014}$ \\
$E_{\rm cut}$ [keV] & $193_{-44}^{+57}$ & $142_{-28}^{+50}$ & $193_{-51}^{+92}$ & $500_{-183}$
& 
$344_{-109}^{\rm +(P)}$ & $183_{-39}^{+61}$ & $162_{-35}^{+46}$ & $201_{-47}^{+84}$ \\
$N_\text{\sc cutoffpl}$~$(10^{-3})$ & $8.8_{-0.5}^{+0.4}$ & $13.24_{-0.5}^{+0.3}$ & $19.7_{-1.3}^{+0.8}$ & $27.3_{-1.7}^{+1.9}$ 
& $8.7_{-0.4}^{+0.4}$ & $12.9_{-0.7}^{+0.5}$ & $17.8_{-1.2}^{+1.0}$ & $23.9_{-1.8}^{+1.6}$ \\ 
\hline
{\sc relconv} &&&& \\
$q_{\rm in}$ & $2.91_{-0.16}^{+0.18}$ & $2.87_{-0.13}^{+0.19}$ & $2.86_{-0.22}^{+0.21}$ & $2.9_{-0.3}^{+0.3}$
& $0.9_{\rm -(P)}^{+3.2}$ & $3.9_{-2.5}^{+1.4}$ & $4.1_{-0.5}^{+0.6}$ & $10.0_{-1.7}$ \\
$q_{\rm out}$ & $= q_{\rm in}$ & $= q_{\rm in}$ & $= q_{\rm in}$ & $= q_{\rm in}$ 
& $2.86_{-0.35}^{+0.15}$ & $2.6_{-0.7}^{+0.4}$ & $1.6_{-0.5}^{+0.8}$ & $2.7_{-0.3}^{+0.3}$ \\
$R_{\rm br}$ [$M$] & -- & -- & -- & --
& $3_{\rm -(P)}^{+7}$ & $7_{\rm -(P)}^{\rm +(P)}$ & $19_{-7}^{\rm +(P)}$ & $3.6_{-0.4}^{+3.1}$ \\
$i$ [deg] & \multicolumn{4}{c}{$29.0_{-2.3}^{+2.2}$} 
& \multicolumn{4}{c}{$30.7_{-2.5}^{+2.2}$} \\
$a_*$ & \multicolumn{4}{c}{$0.90_{-0.07}^{+0.06}$} 
& \multicolumn{4}{c}{$0.915_{-0.030}^{+0.024}$} \\
\hline
{\sc xillver} &&&& \\
$z$ & \multicolumn{4}{c}{$0.007749^\star$} & \multicolumn{4}{c}{$0.007749^\star$} \\
$\log\xi$ & $2.85_{-0.04}^{+0.13}$ & $2.87_{-0.04}^{+0.14}$ & $2.94_{-0.12}^{+0.15}$ & $3.056_{-0.056}^{+0.021}$ 
& $2.92_{-0.11}^{+0.09}$ & $2.93_{-0.09}^{+0.08}$ & $3.010_{-0.151}^{+0.023}$ & $3.04_{-0.03}^{+0.04}$ \\
$A_{\rm Fe}$ & \multicolumn{4}{c}{$3.0_{-0.6}^{+0.6}$} 
& \multicolumn{4}{c}{$3.0_{-0.4}^{+0.5}$} \\
$N_\text{\sc xillver}$~$(10^{-3})$ & $0.088_{-0.008}^{+0.008}$ & $0.104_{-0.010}^{+0.009}$ & $0.136_{-0.015}^{+0.017}$ & $0.141_{-0.021}^{+0.021}$ 
& $0.108_{-0.010}^{+0.009}$ & $0.128_{-0.013}^{+0.012}$ & $0.183_{-0.016}^{+0.018}$ & $0.25_{-0.03}^{+0.03}$ \\ 
\hline
{\sc xillver} &&&& \\
$\log\xi'$ & \multicolumn{4}{c}{$0^\star$} & \multicolumn{4}{c}{$0^\star$} \\
$N'_\text{\sc xillver}$~$(10^{-3})$ & \multicolumn{4}{c}{$0.059_{-0.009}^{+0.010}$} 
& \multicolumn{4}{c}{$0.054_{-0.010}^{+0.011}$} \\
\hline
{\sc zgauss} &&&& &&&&\\
$E_{\rm line}$ [keV] & \multicolumn{4}{c}{$0.8237_{-0.0004}^{+0.0018}$} 
& \multicolumn{4}{c}{$0.8050_{-0.0021}^{+0.0022}$} \\
\hline
{\sc zgauss} &&&& &&&&\\
$E_{\rm line}$ [keV] & \multicolumn{4}{c}{$1.241_{-0.010}^{+0.010}$} 
& \multicolumn{4}{c}{$1.224_{-0.012}^{+0.011}$} \\
\hline
$\chi^2$/dof & \multicolumn{4}{c}{$3094.68/2689 = 1.15087$} & \multicolumn{4}{c}{$3079.54/2681 = 1.14865$} \\
\hline\hline
\end{tabular}
\vspace{0.2cm}
\caption{MCG--6--30--15 -- Summary of the best-fit values for the models {\sc relconv$\times$xillver}~(s) and {\sc relconv$\times$xillver}~(b). The reported uncertainties correspond to the 90\% confidence level for one relevant parameter ($\Delta\chi^2 = 2.71$). $\xi$, $\xi'$, $\xi_1$, and $\xi_2$ in units of erg~cm~s$^{-1}$. See the text for the details. \label{t-mcg2}}
}
\end{table*}

\begin{table*}
\centering
{\renewcommand{\arraystretch}{1.25}
\begin{tabular}{l|cccc|cccc}
\hline\hline
Model & \multicolumn{4}{c}{{\sc relxill} (s)} & \multicolumn{4}{c}{{\sc relxill} (b)} \\
\hline
Flux state & 1 & 2 & 3 & 4 & 1 & 2 & 3 & 4 \\
\hline
{\sc tbabs} &&&& &&&& \\
$N_{\rm H} / 10^{22}$ cm$^{-2}$ & \multicolumn{4}{c}{$0.039^\star$} & \multicolumn{4}{c}{$0.039^\star$} \\
\hline
{\sc warmabs$_1$} &&&& \\
$N_{\rm H \, 1} / 10^{22}$ cm$^{-2}$ & $0.67^{+0.10}_{-0.31}$ & $1.06^{+0.07}_{-0.07}$ & $1.03^{+0.06}_{-0.05}$ & $0.65^{+0.06}_{-0.06}$ 
& $0.49^{+0.29}_{-0.12}$ & $1.02^{+0.06}_{-0.08}$ & $0.95^{+0.05}_{-0.07}$ & $0.58^{+0.05}_{-0.06}$ \\
$\log\xi_1$ & $1.842^{+0.050}_{-0.011}$ & $1.90^{+0.03}_{-0.03}$ & $1.912^{+0.019}_{-0.021}$ & $1.77^{+0.03}_{-0.04}$ 
& $1.844^{+0.086}_{-0.017}$ & $1.92^{+0.04}_{-0.04}$ & $1.92^{+0.03}_{-0.03}$ & $1.77^{+0.04}_{-0.05}$ \\
\hline
{\sc warmabs$_2$} &&&& \\
$N_{\rm H \, 2} / 10^{22}$ cm$^{-2}$ & $0.49^{+0.30}_{-0.11}$ & $0.01^\star$ & $0.48^{+0.19}_{-0.19}$ & $0.32^{+0.06}_{-0.07}$ 
& $0.63^{+0.09}_{-0.15}$ & $0.01^\star$ & $0.59^{+0.24}_{-0.17}$ & $0.32^{+0.07}_{-0.08}$ \\
$\log\xi_2$ & $1.85^{+0.11}_{-0.03}$ & $2.00^\star$ & $3.22^{+0.07}_{-0.11}$ & $2.50^{+0.11}_{-0.11}$ 
& $1.88^{+0.06}_{-0.05}$ & $2.00^\star$ & $3.26^{+0.06}_{-0.08}$ & $2.50^{+0.11}_{-0.12}$ \\
\hline
{\sc dustyabs} &&&& \\
$\log \big( N_{\rm Fe} / 10^{21}$ cm$^{-2} \big)$ & \multicolumn{4}{c}{$17.428^{+0.022}_{-0.039}$} 
& \multicolumn{4}{c}{$17.43^{+0.03}_{-0.04}$} \\
\hline
{\sc cutoffpl} &&&& \\
$\Gamma$ & $1.981^{+0.014}_{-0.014}$ & $1.997^{+0.012}_{-0.013}$ & $2.035^{+0.011}_{-0.013}$ & $2.047^{+0.016}_{-0.011}$ 
& $1.950^{+0.018}_{-0.015}$ & $1.969^{+0.011}_{-0.020}$ & $1.997^{+0.016}_{-0.023}$ & $1.999^{+0.023}_{-0.013}$ \\
$E_{\rm cut}$ [keV] & $340^{+157}_{-116}$ & $183^{+59}_{-41}$ & $158^{+50}_{-39}$ & $212^{+125}_{-49}$
& 
$293^{+134}_{-75}$ & $175^{+44}_{-35}$ & $162^{+37}_{-32}$ & $199^{+65}_{-43}$ \\
$N_\text{\sc cutoffpl}$~$(10^{-3})$ & $8.9^{+0.4}_{-0.5}$ & $13.3^{+0.6}_{-0.6}$ & $18.5^{+0.7}_{-0.7}$ & $26.1^{+1.5}_{-1.3}$ 
& $7.9^{+0.6}_{-1.0}$ & $11.6^{+1.2}_{-1.6}$ & $13.6^{+2.2}_{-2.7}$ & $18^{+4}_{-3}$ \\ 
\hline
{\sc relxill} &&&& \\
$q_{\rm in}$ & $3.01^{+0.17}_{-0.15}$ & $3.03^{+0.17}_{-0.17}$ & $3.26^{+0.25}_{-0.23}$ & $3.6^{+0.3}_{-0.4}$
& $6.1^{\rm + (P)}_{-2.5}$ & $4.6^{+2.4}_{-1.0}$ & $6.4^{+1.7}_{-1.2}$ & $6.8^{+1.4}_{-1.0}$ \\
$q_{\rm out}$ & $= q_{\rm in}$ & $= q_{\rm in}$ & $= q_{\rm in}$ & $= q_{\rm in}$ 
& $2.78^{+0.15}_{-0.15}$ & $2.4^{+0.4}_{-0.8}$ & $2.7^{+0.3}_{-0.3}$ & $2.6^{+0.4}_{-0.5}$ \\
$R_{\rm br}$ [$M$] & -- & -- & -- & --
& $3.3^{+3.6}_{-0.9}$ & $7^{+10}_{-3}$ & $4.4^{+2.0}_{-1.1}$ & $4.8^{+2.5}_{-1.2}$ \\
$i$ [deg] & \multicolumn{4}{c}{$33.3^{+1.6}_{-1.6}$} 
& \multicolumn{4}{c}{$28.7^{+2.1}_{-1.0}$} \\
$a_*$ & \multicolumn{4}{c}{$0.997^{\rm + (P)}_{-0.04}$} 
& \multicolumn{4}{c}{$0.962^{+0.019}_{-0.017}$} \\
$z$ & \multicolumn{4}{c}{$0.007749^\star$} & \multicolumn{4}{c}{$0.007749^\star$} \\
$\log\xi$ & $2.91^{+0.09}_{-0.09}$ & $2.91^{+0.10}_{-0.08}$ & $3.00^{+0.03}_{-0.19}$ & $3.01^{+0.04}_{-0.14}$ 
& $3.00^{+0.03}_{-0.10}$ & $3.01^{+0.03}_{-0.16}$ & $3.07^{+0.06}_{-0.04}$ & $3.13^{+0.08}_{-0.05}$ \\
$A_{\rm Fe}$ & \multicolumn{4}{c}{$3.4^{+0.5}_{-0.5}$} 
& \multicolumn{4}{c}{$2.9^{+0.3}_{-0.3}$} \\
$N_\text{\sc relxill}$~$(10^{-3})$ & $0.048^{+0.004}_{-0.004}$ & $0.055^{+0.004}_{-0.004}$ & $0.079^{+0.007}_{-0.006}$ & $0.096^{+0.010}_{-0.013}$ 
& $0.057^{+0.010}_{-0.010}$ & $0.069^{+0.012}_{-0.010}$ & $0.114^{+0.020}_{-0.008}$ & $0.16^{+0.03}_{-0.03}$ \\ 
\hline
{\sc xillver} &&&& \\
$\log\xi'$ & \multicolumn{4}{c}{$0^\star$} & \multicolumn{4}{c}{$0^\star$} \\
$N'_\text{\sc xillver}$~$(10^{-3})$ & \multicolumn{4}{c}{$0.065^{+0.009}_{-0.005}$} 
& \multicolumn{4}{c}{$0.046^{+0.011}_{-0.010}$} \\
\hline
{\sc zgauss} &&&& &&&&\\
$E_{\rm line}$ [keV] & \multicolumn{4}{c}{$0.8044^{+0.0023}_{-0.0017}$} 
& \multicolumn{4}{c}{$0.806^{+0.003}_{-0.003}$} \\
\hline
{\sc zgauss} &&&& &&&&\\
$E_{\rm line}$ [keV] & \multicolumn{4}{c}{$1.224^{+0.010}_{-0.009}$} 
& \multicolumn{4}{c}{$1.222^{+0.012}_{-0.015}$} \\
\hline
$\chi^2$/dof & \multicolumn{4}{c}{$3081.74/2691 = 1.14520$} & \multicolumn{4}{c}{$3024.46/2683 = 1.12727$} \\
\hline\hline
\end{tabular}
\vspace{0.2cm}
\caption{MCG--6--30--15 -- Summary of the best-fit values for the models {\sc relxill}~(s) and {\sc relxill}~(b). The reported uncertainties correspond to the 90\% confidence level for one relevant parameter ($\Delta\chi^2 = 2.71$). $\xi$, $\xi'$, $\xi_1$, and $\xi_2$ in units of erg~cm~s$^{-1}$. See the text for the details. \label{t-mcg3}}
}
\end{table*}

\begin{figure*}
\begin{center}
\includegraphics[width=5.7cm,trim={1.7cm 0cm 2.6cm 17.5cm},clip]{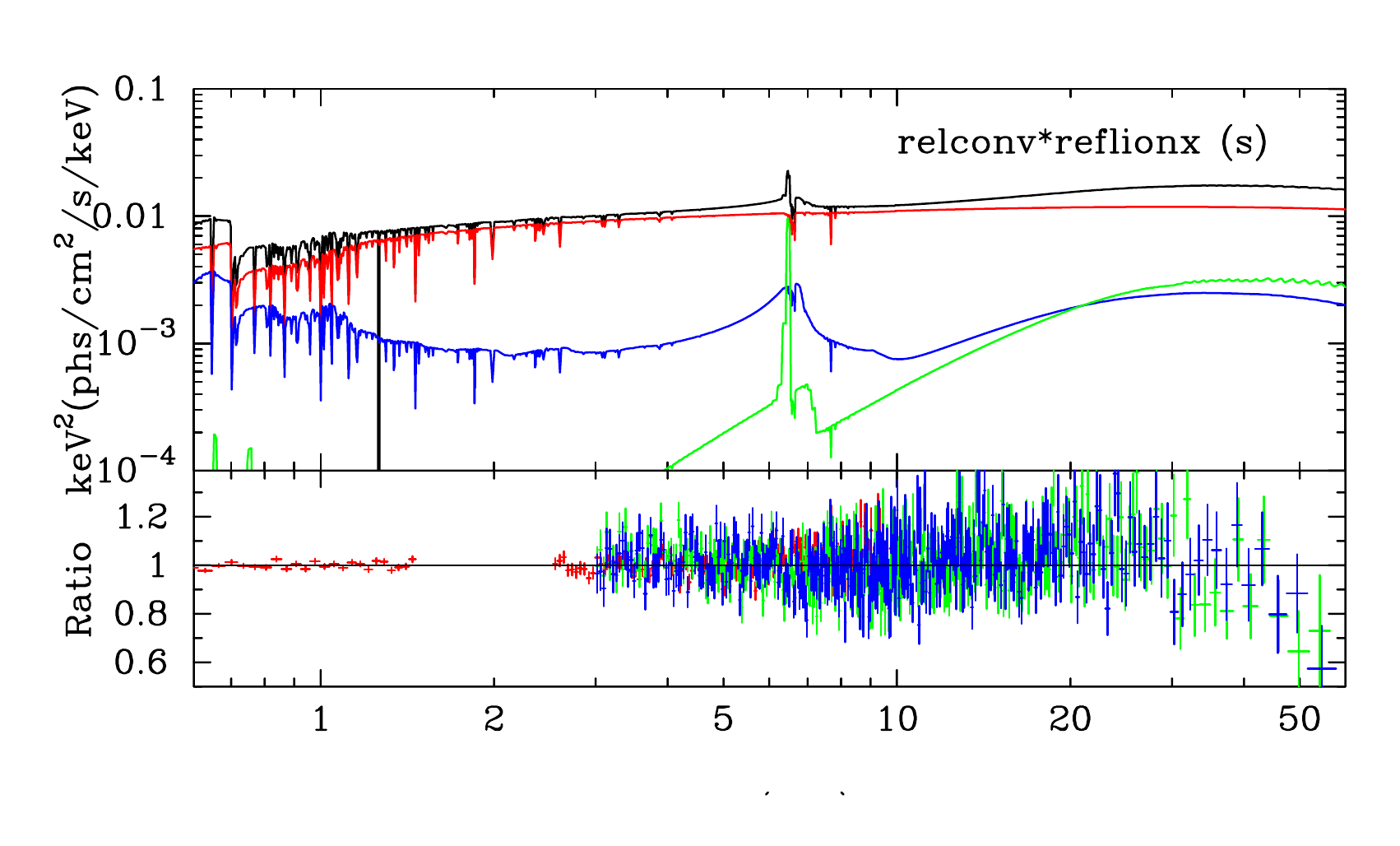}
\hspace{0.0cm}
\includegraphics[width=5.7cm,trim={1.7cm 0cm 2.6cm 17.5cm},clip]{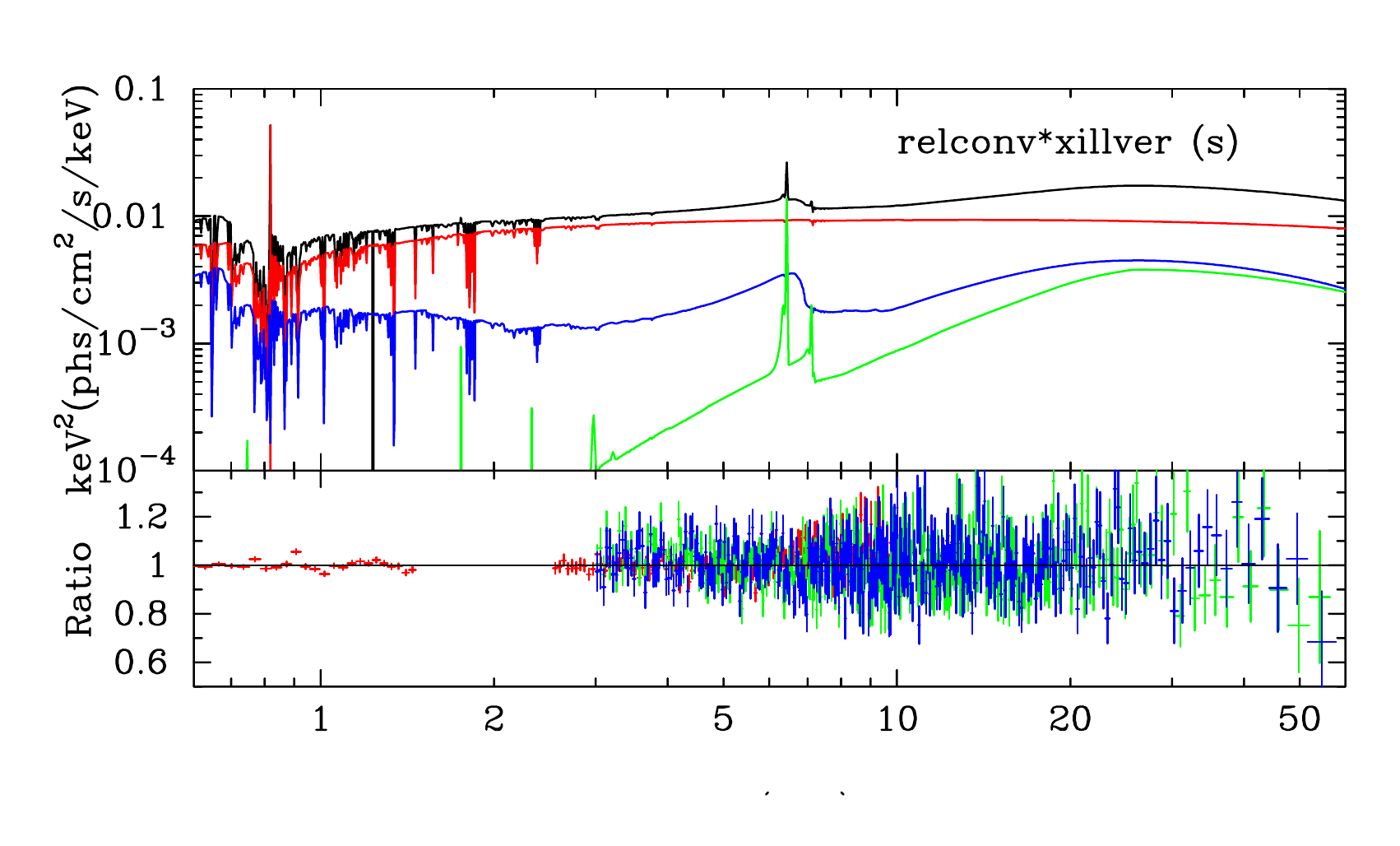}
\hspace{0.0cm}
\includegraphics[width=5.7cm,trim={1.7cm 0cm 2.6cm 17.5cm},clip]{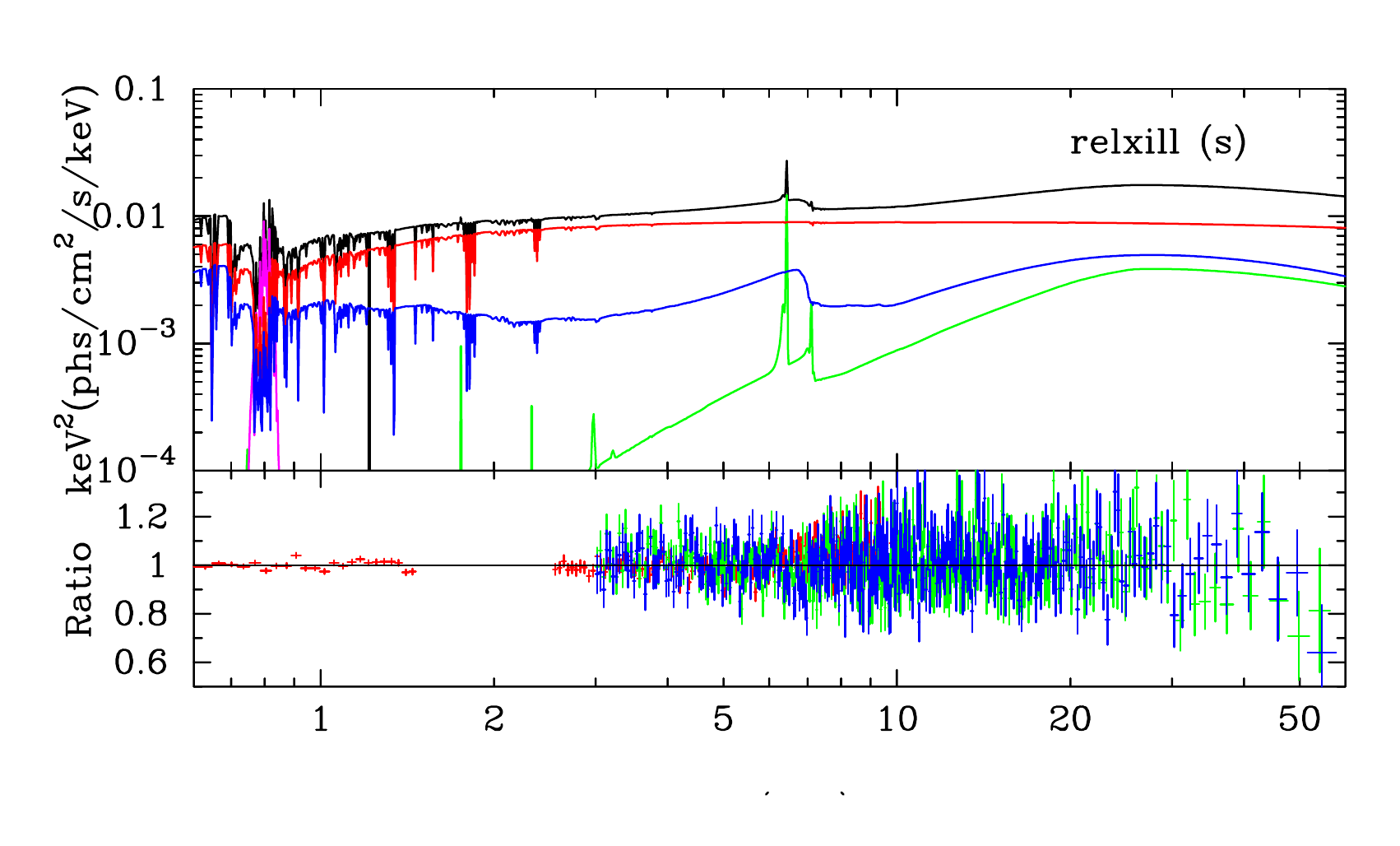} \\
\includegraphics[width=5.7cm,trim={1.7cm 0cm 2.6cm 17.5cm},clip]{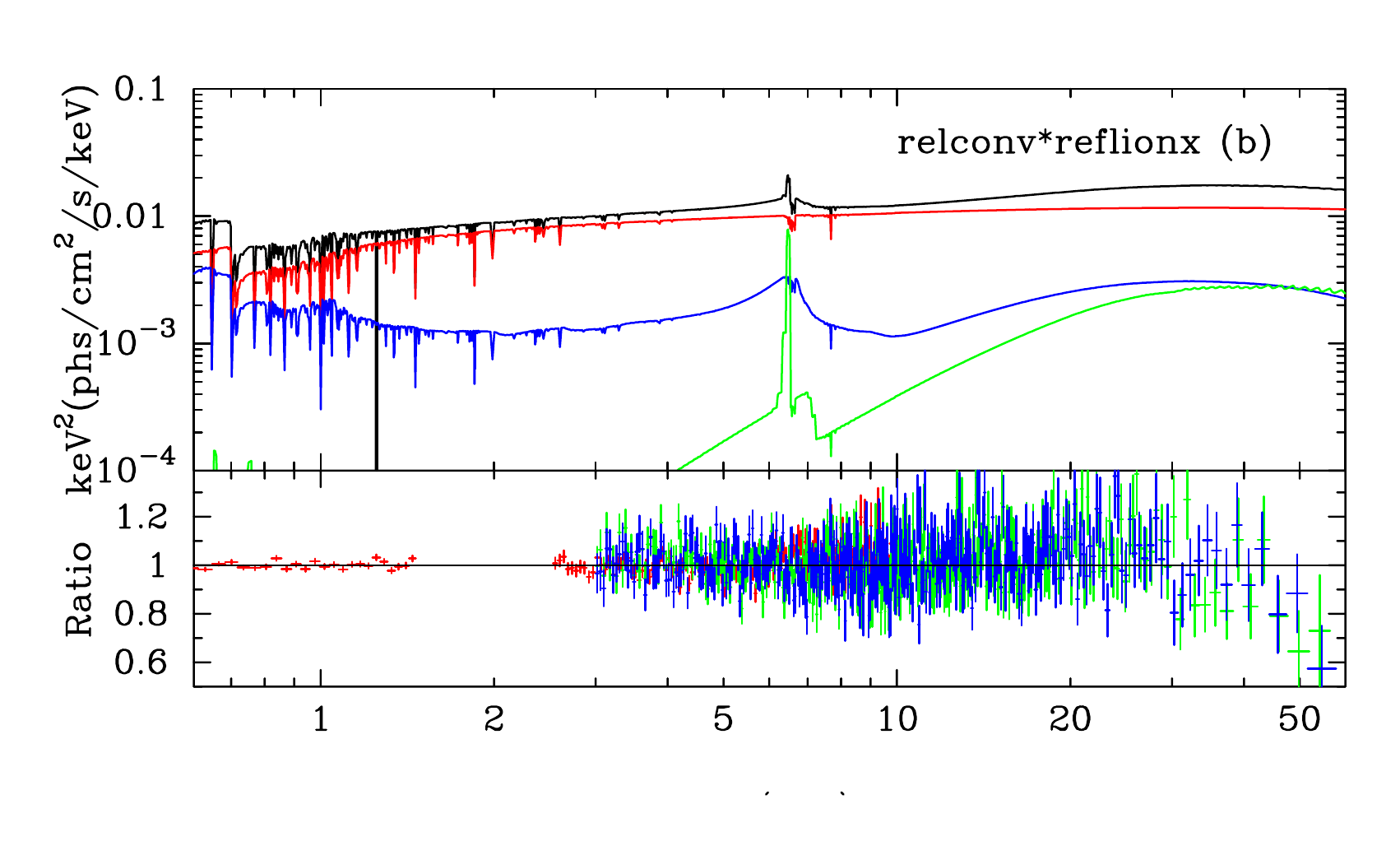}
\hspace{0.0cm}
\includegraphics[width=5.7cm,trim={1.7cm 0cm 2.6cm 17.5cm},clip]{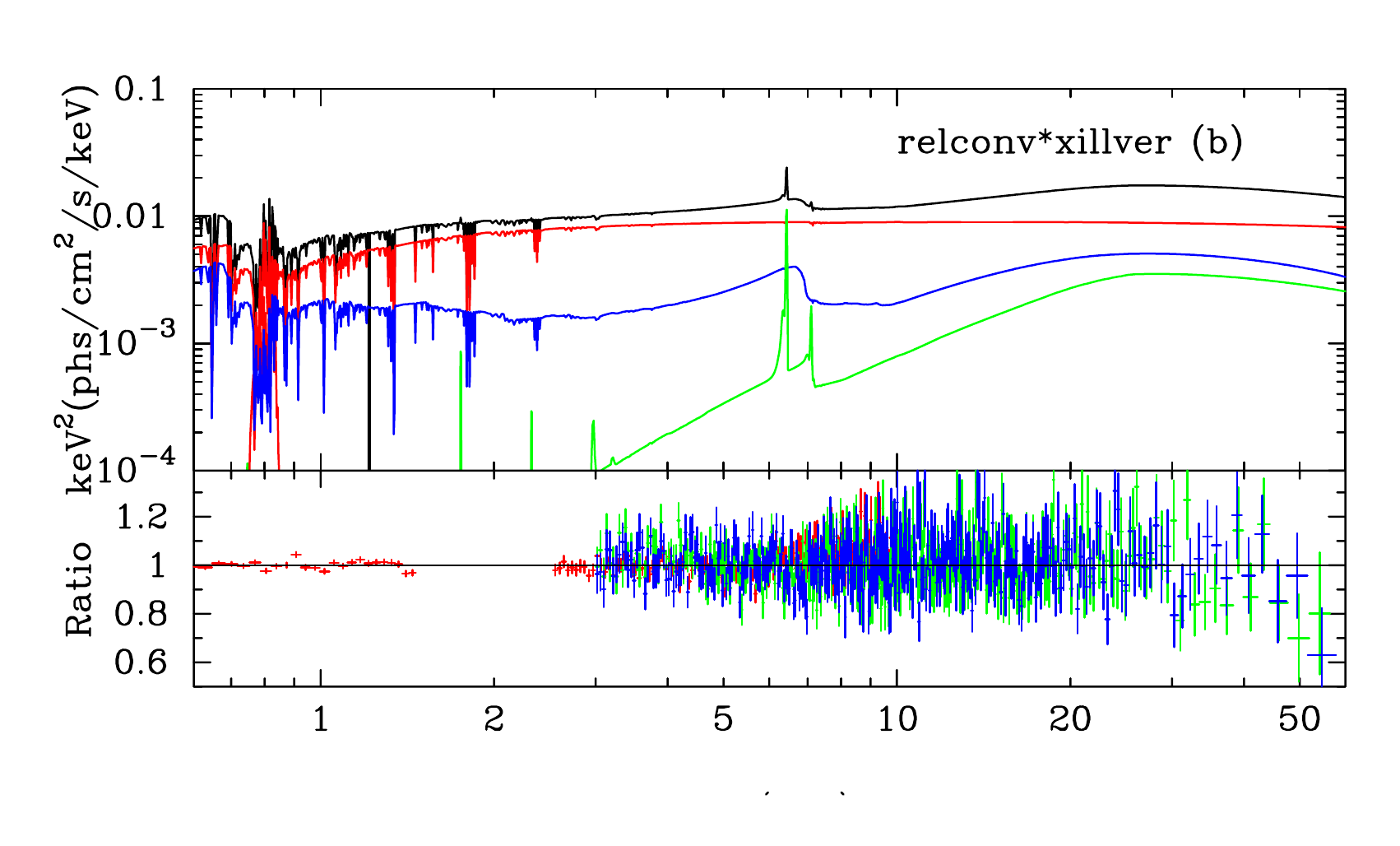}
\hspace{0.0cm}
\includegraphics[width=5.7cm,trim={1.7cm 0cm 2.6cm 17.5cm},clip]{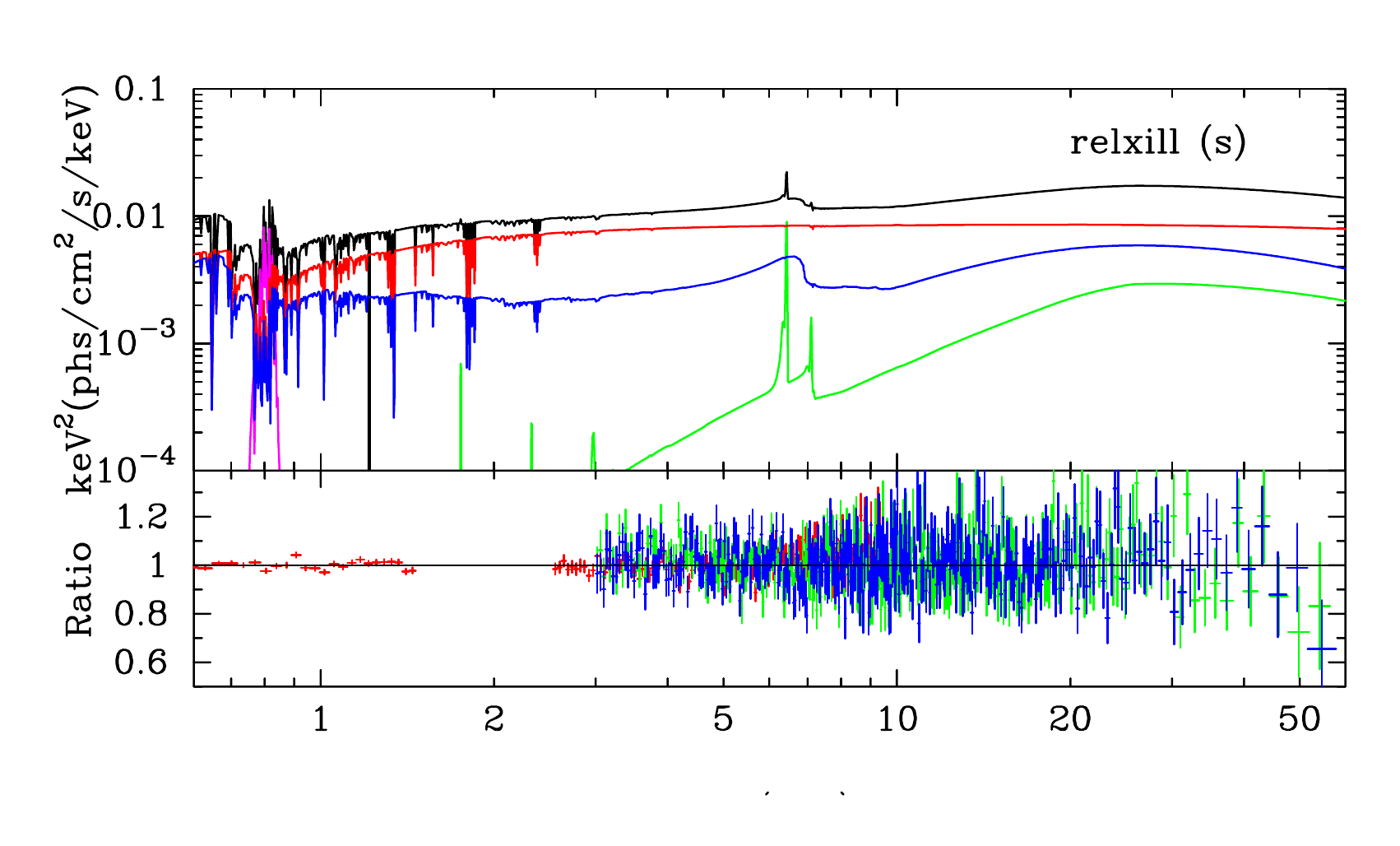} 
\end{center}
\vspace{-0.4cm}
\caption{MCG--6--30--15 -- Best-fit models of the low flux state (top quadrants) and data to best-fit model ratios (bottom quadrants) for the models {\sc relconv$\times$reflionx} (left panels), {\sc relconv$\times$xillver} (central panels), and {\sc relxill} (right panels), employing an emissivity profile described by a simple power-law (top panels) and a broken power-law (bottom panels). In every panel, the total spectrum is in black, the power-law component from the corona is in red, the relativistic reflection component from the disk is in blue, and the non-relativistic reflection component from cold material is in green. \label{f-mcg-mr}}
\vspace{0.2cm}
\begin{center}
\includegraphics[width=5.7cm,trim={0cm 0cm 1.0cm 0.2cm},clip]{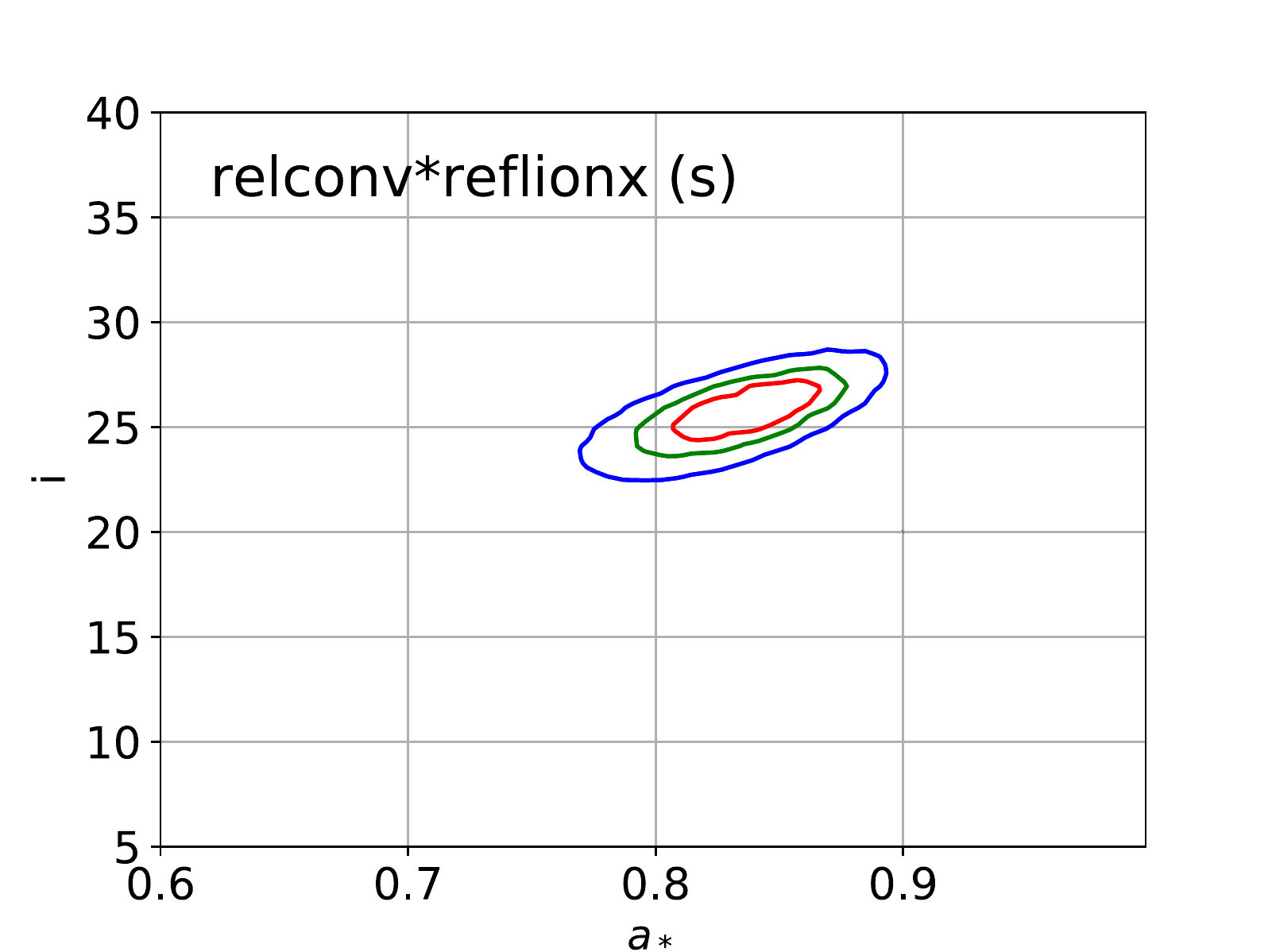}
\hspace{0.0cm}
\includegraphics[width=5.7cm,trim={0cm 0cm 1.0cm 0.2cm},clip]{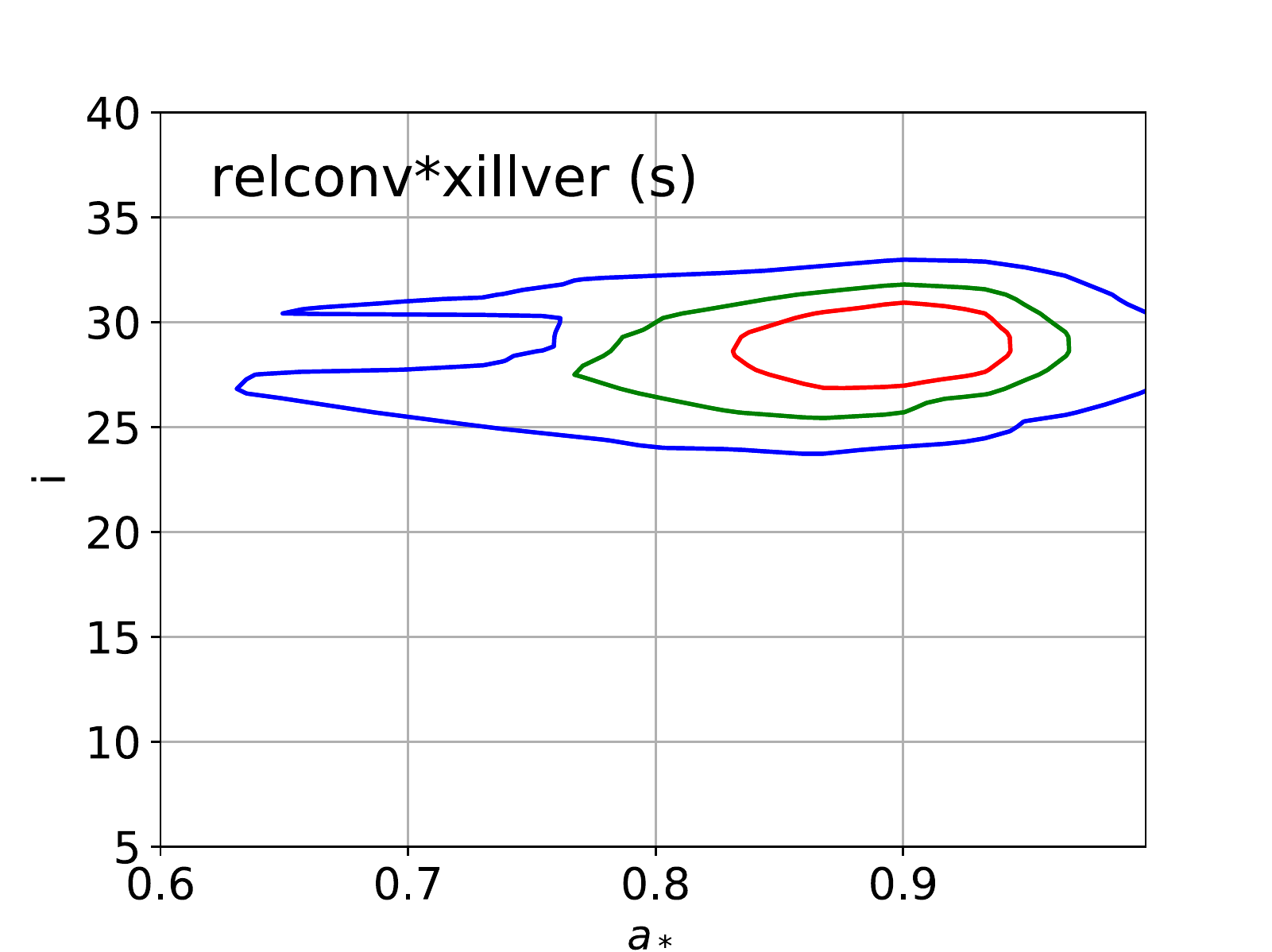}
\hspace{0.0cm}
\includegraphics[width=5.7cm,trim={0cm 0cm 1.0cm 0.2cm},clip]{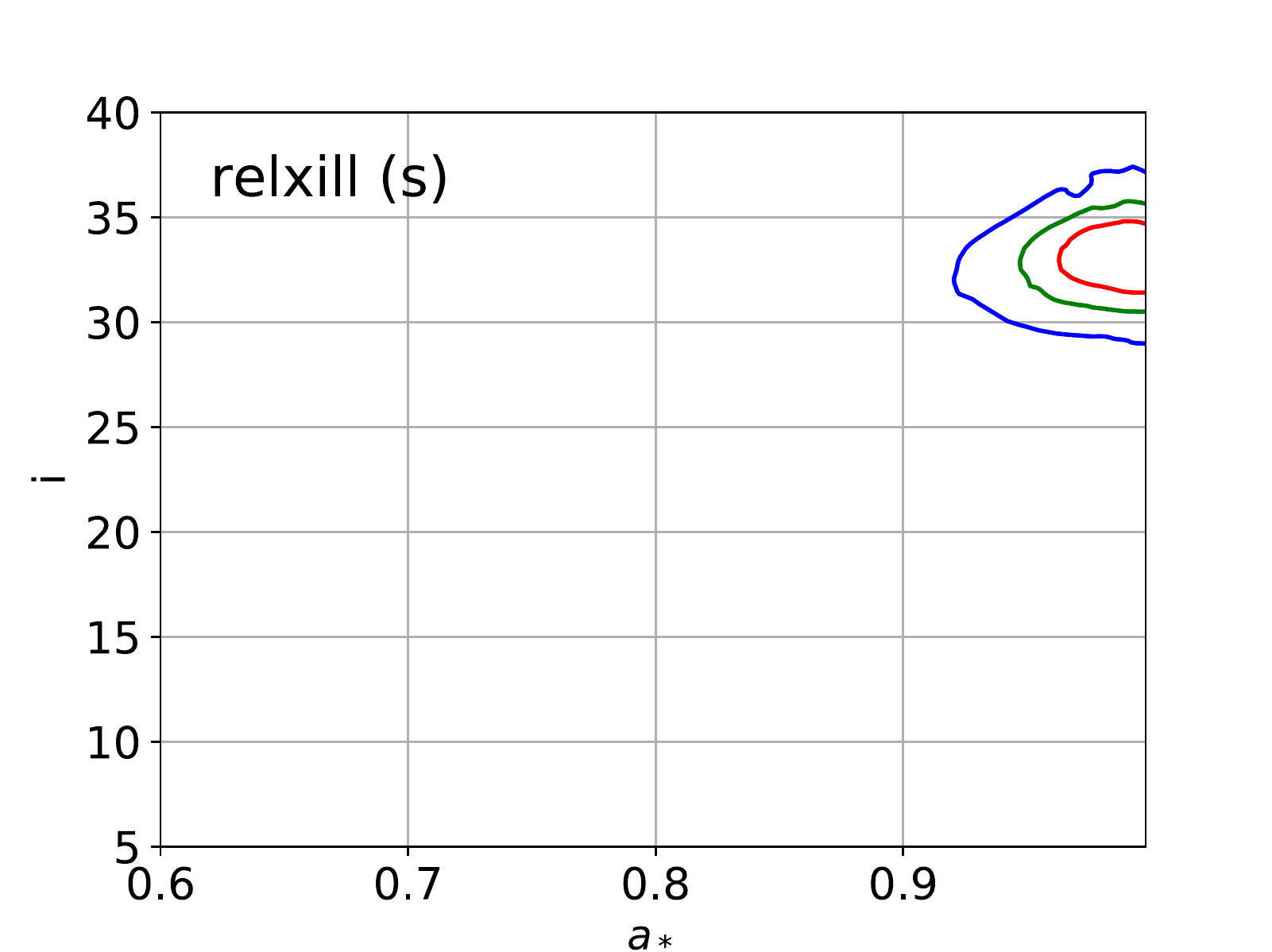} \\
\includegraphics[width=5.7cm,trim={0cm 0cm 1.0cm 0.2cm},clip]{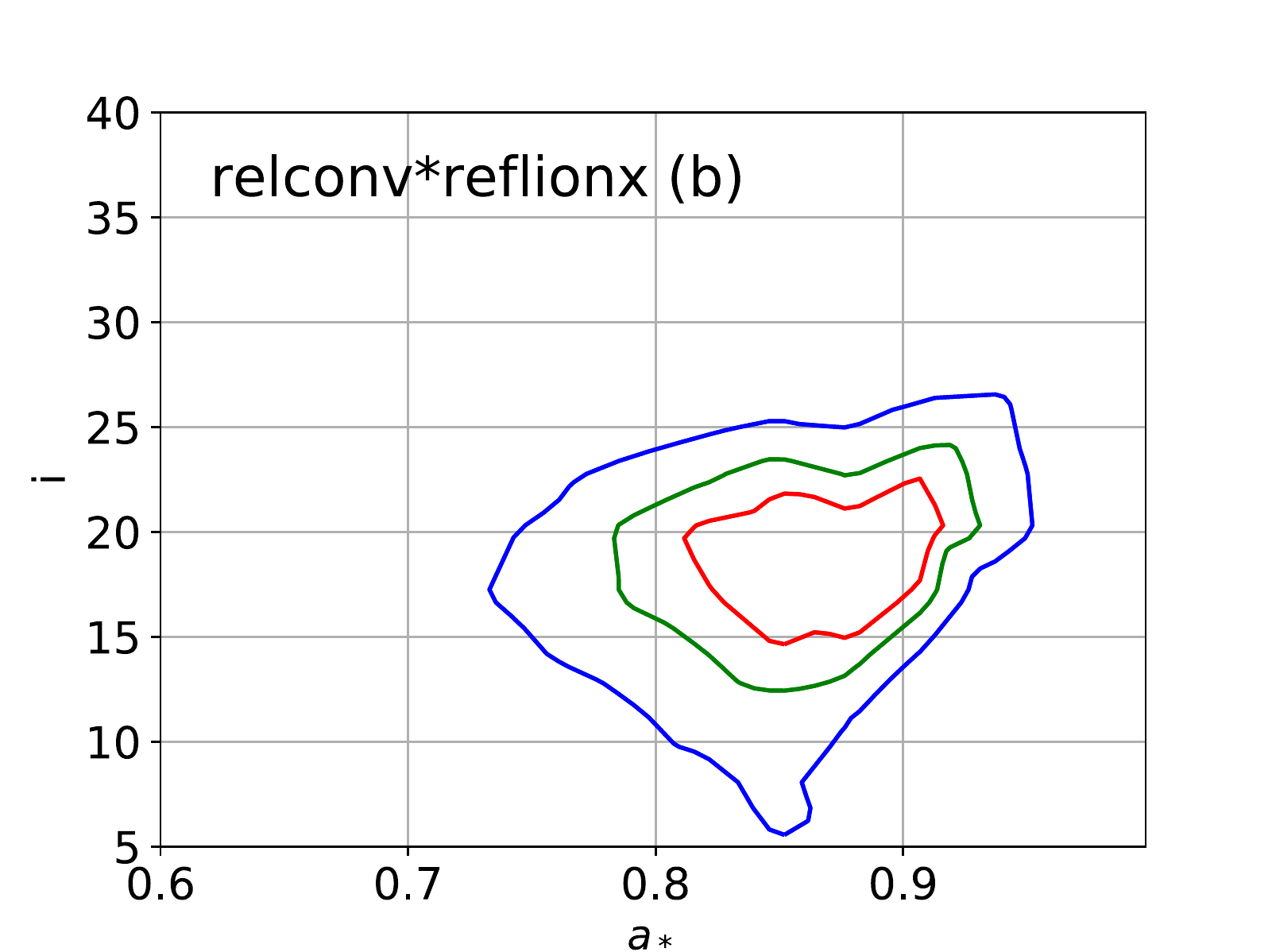}
\hspace{0.0cm}
\includegraphics[width=5.7cm,trim={0cm 0cm 1.0cm 0.2cm},clip]{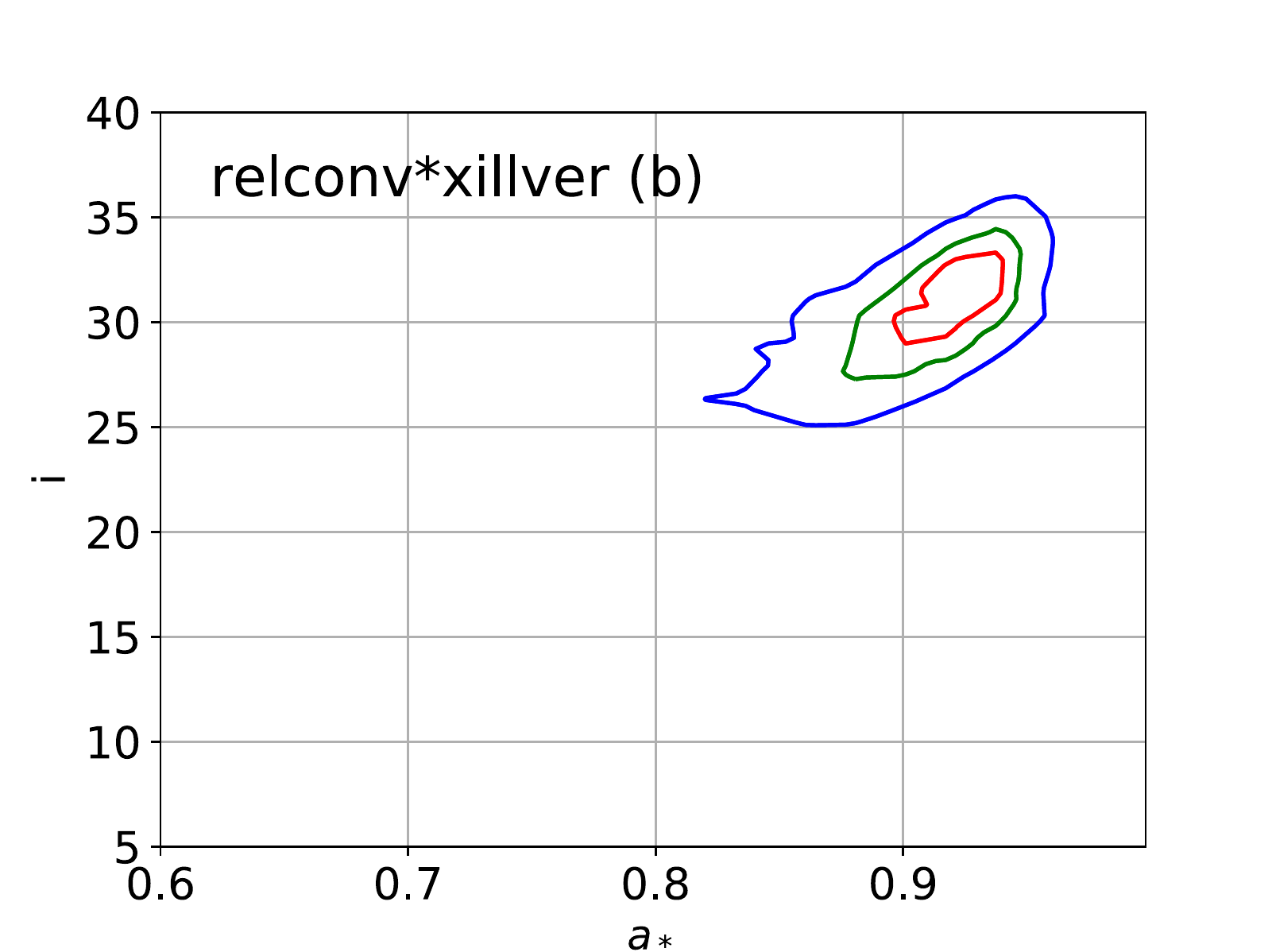}
\hspace{0.0cm}
\includegraphics[width=5.7cm,trim={0cm 0cm 1.0cm 0.2cm},clip]{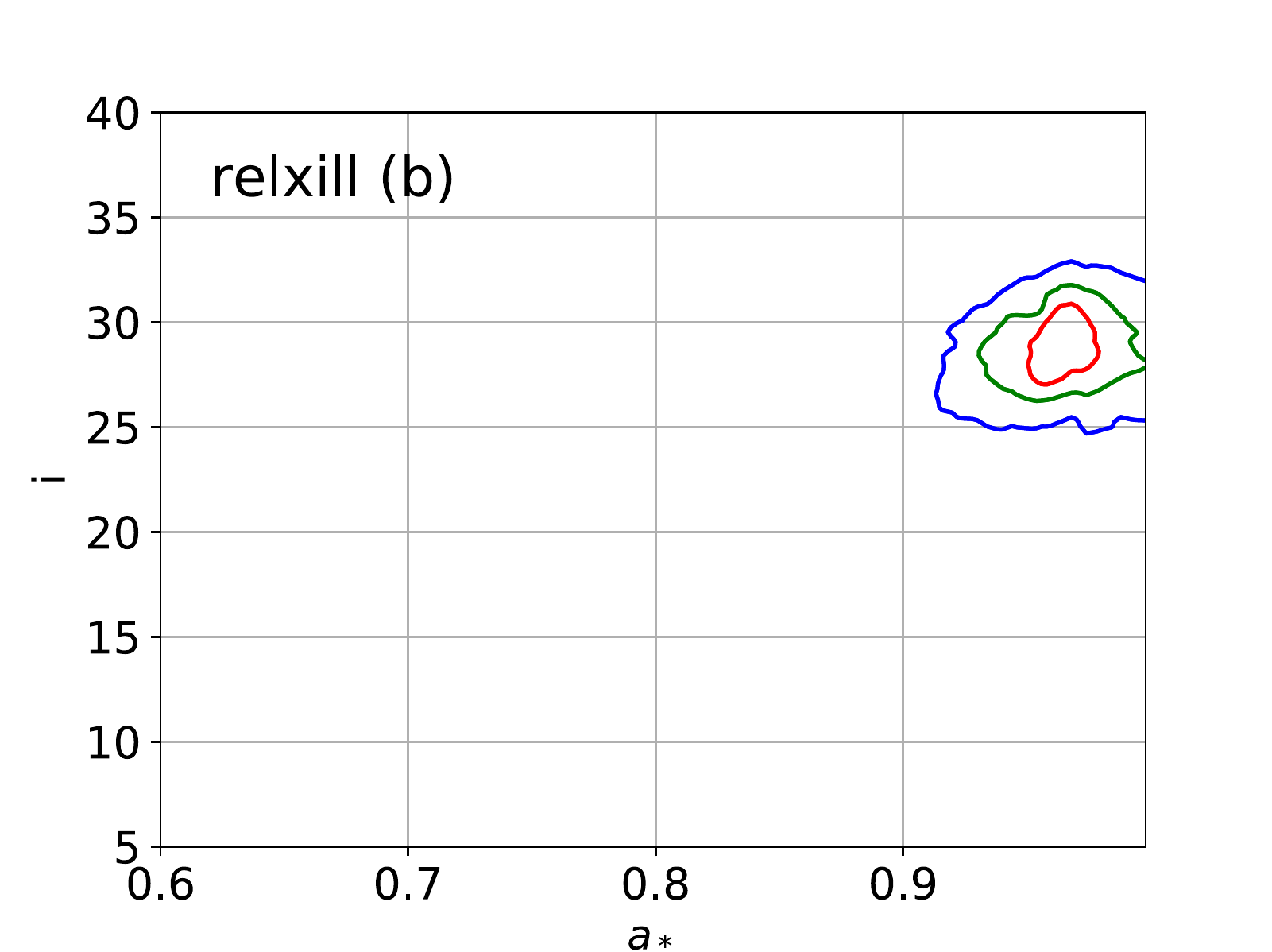}
\end{center}
\vspace{-0.2cm}
\caption{MCG--6--30--15 -- Constraints on the black hole spin parameter $a_*$ and the inclination angle of the disk $i$ for the models {\sc relconv$\times$reflionx} (left panels), {\sc relconv$\times$xillver} (central panels), and {\sc relxill} (right panels), employing an emissivity profile described by a simple power-law (top panels) and a broken power-law (bottom panels). The red, greed, and blue curves correspond, respectively, to the 68\%, 90\%, and 99\% confidence level limits for two relevant parameters. \label{f-mcg-ai}}
\end{figure*}


\section{Discussion and conclusions}\label{s-dc}

Previous studies have already investigated the performance of the reflection model {\sc relxill} in the analysis of X-ray reflection spectra of accreting black holes. In general, it is found that reliable spin measurements can only be obtained if the corona illuminates well the inner part of the accretion disk (within the lamppost set-up, this requires that the corona height is low) and if the spin parameter is high~\citep[see][]{2016AN....337..423B, 2017ApJ...851...57C, 2018A&A...614A..44K, 2019A&A...628A...5B}. The present work is focused on the differences between the reflection models {\sc relconv$\times$reflionx}, {\sc relconv$\times$xillver}, and {\sc relxill} and we have analyzed a \textsl{Suzaku} observation of the high viewing angle black hole binary GRS~1915+105 and simultaneous \textsl{XMM-Newton} and \textsl{NuSTAR} observations of the low viewing angle AGN MCG--6--30--15. Both sources are usually thought to host a fast rotating black hole [GRS~1915+105:~\citet{2013ApJ...775L..45M}, MCG--6--30--15:~\citet{2014ApJ...787...83M}]. The steep emissivity of GRS~1915+105 also implies a corona that illuminates well the inner part of the disk, which is a required ingredient for a reliable measurement.

For the \textsl{Suzaku} observation of GRS~1915+105, the three models provide quite similar fits. The estimates of the model parameters are consistent, with only some minor discrepancy in the estimate of the high energy cut-off $E_{\rm cut}$ of the coronal spectrum. We find a high inclination angle ($i\approx 70^{\circ}$) is always needed to fit the data, which is consistent with previous studies with \textsl{NuSTAR} observations~\citep[e.g.][]{2013ApJ...775L..45M, 2020MNRAS.492..405S}. All models require a very high inner emissivity index $q_{\rm in}$ and an outer emissivity index $q_{\rm out}$ close to zero. 
Such an emissivity profile was already found for GRS~1915+105 in a \textsl{NuSTAR} observation of 2012~\citep{2013ApJ...775L..45M}, as well as for other black hole binaries, such as in the 2015 \textsl{NuSTAR} observation of GS~1354--645 analyzed in~\citet{2016ApJ...826L..12E}. A similar emissivity profile can be explained by a ring-like or disk-like corona above the accretion disk and with the axis parallel to the black hole spin~\citep{2003MNRAS.344L..22M,2011MNRAS.414.1269W,2015MNRAS.449..129W}. As shown in Fig.~\ref{f-grs-qi}, there is a correlation between $q_{\rm in}$ and $i$. Such a high value for $q_{\rm in}$ imposes a strong gravitational redshift to the iron line, and a way to compensate such an extreme redshift is a high viewing angle $i$, permitting a strong Doppler blueshift and moving the blue side of the iron line to higher energies. A similar effect was reported in \citet{2018ApJ...864...25G} for the black hole binary in XTE~J1752--223; in that case, a lamppost model was able to reproduce the data with a significantly lower viewing angle, and we may wonder whether a lamppost profile may reduce the value of the viewing angle and provide a better fit even here for GRS~1915+105. If we replace {\sc relxill} with {\sc relxilllp} and we refit the \textsl{Suzaku} data of GRS~1915+105, we find a very low viewing angle $i \approx 30^\circ$, which has never been reported for this source, a higher $\chi^2$ ($\Delta\chi^2 \approx 65$ with respect to the broken power law profile), and the spin cannot be constrained. The lamppost set-up does not seem the correct model for this observation and we prefer the previous interpretation of a ring-like or disk-like corona.

Another issue concerns the possibility of the presence of a distant reflector, as we see some unresolved features in the ratio plots in Fig.~\ref{f-grs-mr} and a distant reflector was reported previously for this source \citep[e.g.][]{2020MNRAS.492..405S}. However, if we add a non-relativistic reflection component to the \textsl{Suzaku} observation, the fit only improves marginally: $\chi^2$ is only a bit lower (for example, in the model with {\sc relxill}, we find $\Delta\chi^2 \approx -3$), the normalization of the non-relativistic reflection component is very low, and the unresolved features in the ratio plot remain without significant changes. Once again, we are instead tempted to argue that the interpretation of the ring-like corona is more convincing. Indeed, in the presence of a ring-like corona above the accretion disk, we should expect the Comptonization of the relativistic reflection component. When such a reflection Comptonization is not included in the model, we find residuals similar to those appearing in Fig.~\ref{f-grs-mr}~\citep{2015MNRAS.448..703W}.

The fact that we obtain very similar results with {\sc relconv$\times$xillver} and {\sc relxill} suggests that the angle dependence of the spectrum emitted from every point of the accretion disk does not play any appreciable role in the \textsl{Suzaku} data of GRS~1915+105. Since we find similar estimates of the model parameters from {\sc relconv$\times$reflionx} and {\sc relconv$\times$xillver}, we conclude that even the choice of the reflection model does not affect the estimate of the properties of the system. Since {\sc reflionx} and {\sc xillver} have been developed independently by different groups, we would be tempted to conclude that systematic uncertainties in the calculation of the reflection spectrum in the rest-frame of the gas in the disk (which are presumably different in {\sc reflionx} and {\sc xillver}) are under control and negligible for the quality of the available data.

The case of MCG--6--30--15 is a bit different. The spectrum of the source is more complicated and there are more components, which makes the comparison among different models not as direct as in the case of GRS~1915+105. For every relativistic reflection model, we show both the fit with the disk emissivity profile modeled by a simple power-law and by a broken power-law. For example, when we use {\sc relconv$\times$xillver} the quality of the two fits is quite similar ($\Delta\chi^2 = 15$), while for {\sc relxill} there is more difference ($\Delta\chi^2 = 57$). In general, the choice of the emissivity profile is thought to play an important role and can affect the estimate of the model parameters; see, for example, \citet{2019PhRvD..99l3007L,2019ApJ...884..147Z}. However, if we see the best-fit tables, and in particular Fig.~\ref{f-grs-ai}, we can realize that here the choice of the relativistic reflection model is more important than the choice between simple and broken power-law. Note that using different relativistic reflection models may affect also the choice of the total model. In particular, when we employ {\sc relconv$\times$reflionx} we do not need {\sc zgauss} to describe the emission line at 0.8~keV (see Tab.~\ref{t-mcg1}). When we use {\sc relxill}, we do not need the second warm absorber in the medium flux state ($N_{\rm H \, 2}$ is frozen to the minimum value 0.01 in Tab.~\ref{t-mcg3}).

The differences between the fits of {\sc relconv$\times$reflionx} and {\sc relconv$\times$xillver} arise from the different reflection model, namely {\sc reflionx} or {\sc xillver}. 
Note that the differences between {\sc reflionx} are {\sc xillver} are larger for intermediate levels of ionization, say $2 < \log\xi < 3$, which is the case of the accretion disk of MCG--6--30--15, while for lower or higher values of $\log\xi$ the differences in the predicted spectra are smaller~\citep{2013ApJ...768..146G}.
We see that the values of the black hole spin parameter $a_*$ and the inclination angle of the disk $i$ inferred with {\sc reflionx} are slightly lower than the values found with {\sc xillver}. Employing a simple power-law profile for the disk emission, the ionization of the accretion disk decreases as the flux of the source increases with {\sc reflionx}, while we see a ionization parameter that increases as the flux of the source increases with {\sc xillver}. In the case of a broken power-law, both models require a relatively constant ionization parameter. While all these trends are possible, they require different coronal geometries.

From the comparison of the fits of {\sc relconv$\times$xillver} and {\sc relxill} we can evaluate the impact of the proper calculations of the angle dependence of the reflection spectrum at the emission point. The angle-resolved model was introduced in \citet{2014ApJ...782...76G}, where the authors showed that angle-averaged models can overestimate the iron abundance by up to a factor of 2. Moreover, in the case of a steep emissivity and high ionization, a high spin will be underestimated if viewed at low inclinations ($\theta<50^{\circ}$) and modeled with angle-averaged models. \citet{2016ApJ...821...11L} point out this issue about angle-averaged and angle-resolved models in the discussion of different spin measurements of Fairall~9.

Lastly, we want to estimate the impact of the choice of the reflection model in the case of future observations of a source like GRS~1915+105, as in our analysis we do not see an appreciable difference among the three models employed. To address this point, we simulate a 20~ks observation with the X-IFU instrument on board of \textsl{Athena}~\citep{2013arXiv1306.2307N}. \citet{2019A&A...628A...5B} conducted comprehensive simulations with the model {\sc relxill} and found that the instrument X-IFU is capable to accurately recover the input parameters. The input model for our simulation is {\sc tbabs$\times$(cutoffpl + relxill)} and the input parameters are the best-fit values in Tab.~\ref{t-grs} found with {\sc relxill}. The observation is generated with the \verb6fakeit6 command in XSPEC and then analyzed with the three models already used in Section~\ref{s-grs}. The results of our fits are shown in Tab.~\ref{t-simu}. {\sc relconv$\times$xillver} and {\sc relxill} provide very similar results, and the fits are good. We can thus conclude that the difference between the best-fit parameters of GRS~1915+105 using the angle-averaged and angle-resolved models will remain negligible even for \textsl{Athena}. However, the choice of the reflection model will be somewhat important. With {\sc relconv$\times$reflionx}, the fit is not good, as $\chi^2/\nu = 1.7$, which means the differences in the calculations of {\sc reflionx} and {\sc xillver} can be clearly detected with \textsl{Athena}. This should not be surprising, as even here we have $2 < \log\xi < 3$, which is the ionization range for which the differences in the predicted spectra of {\sc reflionx} are {\sc xillver} are larger~\citep{2013ApJ...768..146G}. Despite that, it is quite remarkable that the measurements of most parameters of the system are consistent. A clear discrepancy is only found in the estimate of $E_{\rm cut}$. We have some minor discrepancy even in the estimates of $a_*$ and $i$, but the differences are small and we may expect that systematic uncertainties in the description of the accretion disk and the corona are large enough to make the measurements of $a_*$ and $i$ from the three models still consistent.

In conclusion, in this work we have shown that the choice of the relativistic reflection model may have an impact on the estimate of some properties of an accreting black hole already with the available X-ray data. We have employed the most commonly used reflection models among the X-ray astronomy community, namely {\sc reflionx}, {\sc xillver}, and the angle-resolved model {\sc relxill}. Because of the complexity of these systems, it is not straightforward to generalize our results for generic sources and X-ray missions. The particular choice of the source, of its spectral state, of the total XSPEC model, instrumental effects, etc. can combine together in a quite complicated way. In our analysis of the \textsl{Suzaku} observation of GRS~1915+105, we have not found discrepancies in the analysis with different models and between angle-averaged and angle resolved models. Simulating a future observation of GRS~1915+105 with \textsl{Athena}, we have found some minor differences between {\sc reflionx} and {\sc xillver}/{\sc relxill}. In the case of the simultaneous \textsl{XMM-Newton} and \textsl{NuSTAR} observations of MCG--6--30--15, we have seen the difference between angle-averaged and angle-resolved models, which shows the importance of taking the exact emission angle into account as already noted in \citet{2014ApJ...782...76G}.

\begin{table*}
    \centering
    \renewcommand\arraystretch{1.25}
    \begin{tabular}{lccc}
        \hline\hline
        Model  & {\sc relconv $\times$ reflionx} & {\sc relconv $\times$ xillver} & {\sc relxill} \\
        \hline
        {\sc Tbabs} & & & \\
        $N_{\rm H} (10^{22} {\rm cm}^{-2})$ & $5.243_{-0.011}^{+0.009}$ & $5.1936_{-0.0011}^{+0.0011}$ & $5.1942_{-0.0013}^{+0.005}$ \\
        \hline
        {\sc cutoffpl} & & & \\
        $\Gamma$ & $2.289_{-0.006}^{+0.005}$ & $2.1973_{-0.0015}^{+0.0007}$ & $2.1974_{-0.0015}^{+0.006}$\\
        $E_{\rm cut}$ (keV) & $>490$ & $68.3_{-0.9}^{+0.7}$ & $68.8_{-2.0}^{+1.0}$ \\
        $N_{\rm cutoffpl}$ & $4.036_{-0.027}^{+0.014}$ & $3.860_{-0.018}^{+0.05}$ & $3.821_{-0.016}^{+0.028}$ \\
        \hline
        $q_{\rm in}$  & $8.63_{-0.06}^{+0.07}$ & $>9.9$ & $9.925_{-0.027}^{+0.05}$\\
        $q_{\rm out}$ & $0.0^{+0.016}$  & $0.0^{+0.05}$ & $0.0^{+0.13}$ \\
        $R_{\rm br}$ & $7.47_{-0.10}^{+0.09}$ & $6.334_{-0.019}^{+0.03}$ & $6.310_{-0.04}^{+0.019}$ \\
        $a_*$ & $0.998_{-0.00005}$ & $0.9910_{-0.0009}^{+0.0002}$ & $0.9910_{-0.0006}^{+0.0003}$\\
        $i$ (deg) & $74.56_{-0.19}^{+0.25}$ & $73.87_{-0.06}^{+0.08}$ & $73.85_{-0.06}^{+0.08}$\\
        $\log\xi$ & $2.760_{-0.003}^{+0.006}$ & $2.763_{-0.005}^{+0.005}$ & $2.769_{-0.010}^{+0.004}$ \\
        $A_{\rm Fe}$ & $0.601_{-0.016}^{+0.007}$ & $0.579_{-0.005}^{+0.004}$ & $0.575_{-0.004}^{+0.007}$ \\
        $N_{\rm reflionx}$ ($10^{-4}$) & $2.15_{-0.11}^{+0.10}$ & $-$ & $-$\\
        $N_{\rm xiller}$ & $-$ & $0.1533_{-0.004}^{+0.0015}$ & $-$\\
        $N_{\rm relxill}$ ($10^{-2}$)& $-$ & $-$ & $2.106_{-0.025}^{+0.013}$\\
        \hline
        $\chi^2/\nu$ & 13729/7888=1.741 & 7960.6/7888=1.009 & 7896.06/7888=1.001 \\
        \hline\hline       
    \end{tabular}
    \caption{Summary of the best-fit values of the 20~ks simulated observation with X-IFU/\textsl{Athena} fitted with the models {\sc relconv$\times$reflionx}, {\sc relconv$\times$xillver}, and {\sc relxill}. The reported uncertainties correspond to the 90\% confidence level for one relevant parameter ($\Delta\chi^2 = 2.71$). $\xi$ in units of erg~cm~s$^{-1}$. In {\sc reflionx}, the fitting parameter is $\xi$, not $\log\xi$ as in {\sc xillver} and {\sc relxill}, but it is converted into $\log\xi$ in the table in order to facilitate the comparison with the other models. See the text for the details. \label{t-simu}}
\end{table*}


\section*{Acknowledgements}

We would like to thank the referee, Javier Garcia, for useful comments and suggestions to improve the manuscript.
This work was supported by the Innovation Program of the Shanghai Municipal Education Commission, Grant No.~2019-01-07-00-07-E00035, and the National Natural Science Foundation of China (NSFC), Grant No.~11973019.
A.T., C.B., and H.L. are members of the International Team~458 at the International Space Science Institute (ISSI), Bern, Switzerland, and acknowledge support from ISSI during the meetings in Bern.








\bsp	
\label{lastpage}
\end{document}